\documentclass[a4paper,twocolumn,accepted=2023-09-04]{quantumarticle}
\pdfoutput=1

\usepackage[bbgreekl]{mathbbol}
\usepackage{amsfonts}
\DeclareSymbolFontAlphabet{\mathbb}{AMSb}
\DeclareSymbolFontAlphabet{\mathbbl}{bbold}

\usepackage{gensymb,amssymb,graphicx,color,amsmath,mathtools,bm}
\usepackage{hyperref}
\usepackage[normalem]{ulem}
\usepackage[caption=false]{subfig} 
\usepackage{braket}
\usepackage{adjustbox} 
\usepackage{stackrel}

\newcommand{\secref}[1]{Sec.\,\ref{#1}}

\newcommand{\eqnref}[1]{Eq.\,\eqref{#1}}
\newcommand{\eqsref}[1]{Eqs.\,\eqref{#1}}
\newcommand{\figref}[1]{Fig.\,\ref{#1}}
\newcommand{\figsref}[1]{Figs.\,\ref{#1}}
\newcommand{\sfigref}[2]{Fig.\,\hyperref[#1]{\ref{#1}#2}}

\newcommand{\appref}[1]{Appendix\,\ref{#1}}

\definecolor{kspink}{RGB}{200,0,200}

\DeclareMathOperator{\tr}{tr}
\newcommand{\bBra}[1]{\big\langle#1\big|}
\newcommand{\bKet}[1]{\big|#1\big\rangle}
\newcommand{\bBraket}[1]{\big\langle #1 \big\rangle}
\newcommand{\floor}[1]{{\lfloor #1 \rfloor}}
\newcommand{\ceil}[1]{{\lceil #1 \rceil}}

\hypersetup{
  colorlinks = true,
  urlcolor = blue,
  pdfauthor = {Kevin Slagle},
  pdftitle = {Slagle - 2023 - Quantum Gauge Networks: A New Kind of Tensor Network}
}

\begin{document}

\title{Quantum Gauge Networks: A New Kind of Tensor Network}

\author{Kevin Slagle}
\affiliation{Department of Electrical and Computer Engineering, Rice University, Houston, Texas 77005 USA}
\affiliation{Department of Physics, California Institute of Technology, Pasadena, California 91125, USA}
\affiliation{Institute for Quantum Information and Matter and Walter Burke Institute for Theoretical Physics, California Institute of Technology, Pasadena, California 91125, USA}
\orcid{0000-0002-8036-3447}

\begin{abstract}
Although tensor networks are powerful tools for simulating low-dimensional quantum physics,
  tensor network algorithms are very computationally costly in higher spatial dimensions.
We introduce \emph{quantum gauge networks}:
  a different kind of tensor network ansatz for which the computation cost of simulations
  does not explicitly increase for larger spatial dimensions.
We take inspiration from the gauge picture of quantum dynamics \cite{gaugePicture},
  which consists of a local wavefunction for each patch of space,
  with neighboring patches related by unitary connections.
A quantum gauge network (QGN) has a similar structure,
  except the Hilbert space dimensions of the local wavefunctions and connections are truncated.
We describe how a QGN can be obtained from a generic wavefunction or matrix product state (MPS).
All $2k$-point correlation functions of any wavefunction for $M$ many operators
  can be encoded exactly by a QGN with bond dimension $O(M^k)$.
In comparison, for just $k=1$, an exponentially larger bond dimension of $2^{M/6}$ is generically required for an MPS of qubits.
We provide a simple QGN algorithm for approximate simulations of quantum dynamics in any spatial dimension.
The approximate dynamics can achieve exact energy conservation for time-independent Hamiltonians,
  and spatial symmetries can also be maintained exactly.
We benchmark the algorithm by simulating the quantum quench of fermionic Hamiltonians in up to three spatial dimensions.
\end{abstract}

{
  \hypersetup{linkcolor=black}
  \tableofcontents
}
\section{Introduction}

Tensor network algorithms \cite{OrusTN,OrusTNlong,ChanMPS,CiracMPS,RanTN,tensorNetworkQI}
  are very useful for simulating strongly-correlated quantum physics.
In one spatial dimension, matrix product state (MPS) algorithms \cite{OrusTNlong,ChanMPS,CiracMPS} are often
  the best available tool for this task.
Although still useful, tensor network algorithms in higher dimensions
  \cite{OrusTNlong,CiracMPS,IsometricTN,2dDMRG,canonicalPEPS,3DIsoTN,MERA,branchingMERA,entanglementRenormalization,MinimalCanonicalForm,treeRG}
  typically suffer from computational costs that scale as a high power of the bond dimension.
Therefore, we are motivated to study a different kind of tensor network ansatz that is more computationally efficient in many spatial dimensions.

\begin{figure}
  \centering
  \includegraphics{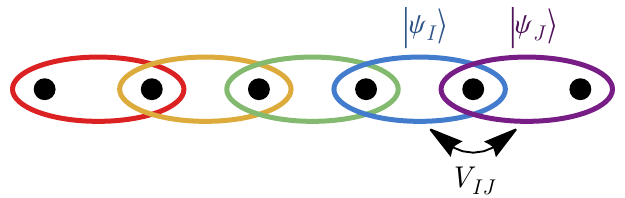}
  \caption{%
    An example of a chain of qubits (black dots) and spatial patches (colored ovals) consisting of pairs of neighboring qubits.
    In the gauge picture,
      a local wavefunction $\ket{\Psi_I}$ is associated with each patch $I$,
      and the Hilbert spaces of neighboring patches are related by unitary connections $\hat{U}_{IJ}$.
    A quantum gauge network analogously consists of local wavefunctions
      $\ket{\psi_I}$ in truncated Hilbert spaces
      that are related by \emph{non-unitary} connections $V_{IJ}$.
  }\label{fig:patches}
\end{figure}

We take inspiration from the gauge picture of quantum dynamics \cite{gaugePicture},
  which adds ``gauge fields'' to Schr\"odinger's picture in order to make spatial locality explicit in the equations of motion.
In the gauge picture,
  one first chooses a collection of possibly-overlapping patches of space that cover space.
For example, one could choose the patches to be pairs of nearest-neighbor sites on a lattice.
We use capital letters, $I$, $J$, or $K$, to denote a spatial patch.
Each patch is assigned a \emph{local wavefunction} $\ket{\Psi_I}$
  (with the same Hilbert space dimension as the usual wavefunction),
  and the Hilbert spaces of neighboring patches are
  related by unitary connections $\hat{U}_{IJ}$ (which act on the entire Hilbert space),
  as depicted in \figref{fig:patches}.
In the simplest setting, the Hamiltonian is written as a sum over terms $\hat{H}_I$ that act within a single patch $I$:
\begin{equation}
  \hat{H} = \sum_I \hat{H}_I \label{eq:H}
\end{equation}
The local wavefunctions and connections time-evolve according to
\begin{equation}
\begin{aligned}
  \partial_t \ket{\Psi_I} &= -i \hat{H}_{\langle I \rangle} \ket{\Psi_I} \\
  \partial_t \hat{U}_{IJ} &= - i \hat{H}_{\langle I \rangle} \hat{U}_{IJ} + i \hat{U}_{IJ} \hat{H}_{\langle J \rangle}
\end{aligned} \label{eq:gaugePicture}
\end{equation}
  where
\begin{equation}
  \hat{H}_{\langle I \rangle} = \sum_J^{J \cap I \neq \emptyset} \hat{U}_{IJ} \, \hat{H}_J \, \hat{U}_{JI} \label{eq:H'hat}
\end{equation}
  is the sum of local Hamiltonian terms supported on patches that overlap with patch $I$.
Typically, we initialize $\hat{U}_{IJ}(0) = \hat{\mathbbl{1}}$ and $\ket{\Psi_I(0)} = \ket{\Psi(0)}$ at time $t=0$,
  where $\hat{\mathbbl{1}}$ is the identity operator and $\ket{\Psi(t)}$ is the usual wavefunction in the Schr\"{o}dinger picture.
The expectation value $\braket{\Psi|\hat{A}_I|\Psi}$ of a local operator $\hat{A}_I$ that only acts within the patch $I$
  can be evaluated in the gauge picture as $\braket{\Psi_I|\hat{A}_I|\Psi_I}$.
To calculate an expectation value $\braket{\Psi|\hat{A}_I \hat{B}_J|\Psi}$
  for a product of operators acting on different patches,
  a connection $\hat{U}_{IJ}$ must be inserted in the gauge picture, as in
  $\braket{\Psi_I| \hat{A}_I \hat{U}_{IJ} \hat{B}_J |\Psi_J}$.
Such correlation functions can be used to calculate the density matrix from the gauge picture local wavefunctions and connections.
For example, for a system of $n$ qubits, the density matrix is
\begin{align}
  \hat{\rho} = 2^{-n} \sum_{\mu_1 \cdots \mu_n} &
    \hat{\sigma}_1^{\mu_1} \cdots \hat{\sigma}_n^{\mu_n} \label{eq:gaugePictureRho}\\
    & \braket{\Psi_1 | \, \hat{\sigma}_1^{\mu_1} \, \hat{U}_{1,2} \, \hat{\sigma}_2^{\mu_2} \cdots \, \hat{U}_{n-1,n} \, \hat{\sigma}_n^{\mu_n} | \Psi_n} \nonumber
\end{align}
  where $\hat{\sigma}_i^{\mu}$ are Pauli operators,
  and here we take each patch to consist of just a single qubit (for simplicity)
  so that $I=1,\ldots,n$ indexes the qubits/patches along some path.
The local wavefunction $\ket{\Psi_I}$ is local in the sense that its dynamics are local and connections are required to extract information about operators outside the patch $I$.
Time evolution preserves the following network of relations:
\begin{equation}
\begin{aligned}
  \hat{U}_{IJ} \ket{\Psi_J} &= \ket{\Psi_I} \\
  \hat{U}_{IJ} \hat{U}_{JK} &= \hat{U}_{IK}
\end{aligned} \label{eq:UIdentities}
\end{equation}
  along with $\hat{U}_{IJ}^\dagger = \hat{U}_{JI}$ and $\hat{U}_{II} = \hat{\mathbbl{1}}$.

In this work, we truncate the Hilbert spaces of the local wavefunctions and connections in the gauage picture
  so that approximate quantum dynamics simulations can be performed on a computer.
The utility of the gauge picture for approximating quantum mechanics
  is that locality is explicit in both the time dynamics and the structure of the local wavefunctions and connections.
Furthermore, the connections allow us to utilize different truncated Hilbert spaces for different patches of space.
We call the resulting network of truncated local wavefunctions and connections a \emph{quantum gauge network} (QGN).

An advantage of quantum gauge networks is that unlike traditional tensor networks (e.g. PEPS \cite{OrusTNlong}),
  a QGN only involves matrices and vectors (rather than tensors with  many indices) regardless of the spatial dimension.
As such, it is natural for a QGN algorithm to only require a computation time (e.g. CPU time) that scales as $O(\chi^3)$, regardless of the number of spatial dimensions,
  where $\chi$ is the dimension of the truncated Hilbert spaces.
$\chi$ can be viewed as the bond dimension of the QGN.
Thus, the natural $O(\chi^3)$ computation time for a QGN algorithm is the same as for MPS algorithms,
  which are very efficient in one spatial dimension (1D).
In three spatial dimensions (3D), the most computationally efficient tensor network
  in previous literature may be the isometric tensor network \cite{IsometricTN,2dDMRG,canonicalPEPS},
  for which the computation time of the TEBD$^3$ algorithm is $O(\chi^{12})$ \cite{3DIsoTN}.\footnote{%
    More generally, the TEBD$^3$ computation time is
      $O(D^{10} \chi^2) + O(D \chi^6)$
      when the bond dimension $\chi$ of the central bonds
      is different from the bond dimension $D$ of the other bonds. \cite{3DIsoTN}
    In 2D, the TEBD$^2$ computation time scales as $\chi^7$. \cite{IsometricTN}}
Thus, QGN algorithms can require significantly
  less computation time for fixed bond dimension.
This is useful since larger bond dimensions allow more correlations to be transported through the tensor network.

The computation time per variational parameter also scales favorable for
  QGN algorithms.
For a QGN, the number of parameters scales as $\chi^2$
  (due to the matrix-valued connections).
Therefore, the computation time per parameter scales as $\chi^3/\chi^2 = \chi^{1.5}$ for a QGN.
This is the same ratio as MPS algorithms,
  which are very efficient in 1D.
For the isometric tensor network TEBD$^3$ algorithm in 3D,
  this ratio scales as $\chi^{12}/\chi^6 = \chi^2$ \cite{3DIsoTN},
  which is remarkably efficient but not as good as the ratio for a QGN or MPS.

Although most tensor networks typically directly encode a wavefunction or density matrix,
  quantum gauge networks depart from this habit.
However, a density matrix can be computed from a QGN similar to \eqnref{eq:gaugePictureRho} for the gauge picture.
Unlike a generic PEPS \cite{OrusTNlong} but similar to a matrix product state (MPS) or isometric tensor network \cite{IsometricTN,2dDMRG,canonicalPEPS},
  local expectation values can be efficiently computed from a QGN.
A disadvantage of quantum gauge networks is that unphysical states can also be encoded.
As such, using a QGN to variationally optimize a ground state is not as straight-forward as for tensor networks that directly encode a wavefunction.
We leave QGN ground state optimization algorithms to future work.
In this work, we focus on QGN fundamentals and time dynamics alorithms.

In \secref{sec:QGN}, we discuss basic properties of quantum gauge networks
  and how a QGN can be constructed.
We also show that for an arbitrary wavefunction (including fermionic wavefunctions),
  all $2k$-point correlation functions of $M$ many operators
  can be encoded exactly by a QGN with bond dimension $O(M^k)$ [\eqnref{eq:OMk}],
  while an MPS of qubits can require an exponentially larger bond dimension $2^{M/6}$ for $k=1$.
In \secref{sec:time}, we present a QGN algorithm for approximately simulating quantum dynamics in the gauge picture.
We benchmark the algorithm using simulations of fermionic Hamiltonians in spatial dimensions up to three.


\section{Quantum Gauge Networks}
\label{sec:QGN}

To define a quantum gauge network (QGN),
  we first choose a collection of possibly-overlapping patches of space that cover space.
For example, one could choose the patches to consist of just a single lattice site.
Another natural choice is to take the patches to have the same support as the Hamiltonian terms.
That is, we might choose patches that are pairs of nearest-neighbor sites if the Hamiltonian terms act on nearest-neighbor sites.
A QGN then consists of
  (1) a \emph{local wavefunction} $\ket{\psi_I}$ for each spatial patch;
  (2) non-unitary \emph{connections} $V_{IJ} = V_{JI}^\dagger$ to relate the Hilbert spaces of nearby patches,
  as depicted in \figref{fig:patches}; and
  (3) a collection of truncated operators to act on the truncated Hilbert space at each patch.

The local wavefunctions and connections are similar to the those within the gauge picture \cite{gaugePicture},
  except the Hilbert space is truncated.
If the full Hilbert space has dimension $N$,
  then $\ket{\Psi_I}$ and $\hat{U}_{IJ}$ in the gauge picture have dimensions $N$ and $N\times N$, respectively.
Since $N$ is exponentially large in system size,
  it is useful to truncate the full Hilbert space dimension for approximate simulations.
Therefore, we consider truncated local wavefunctions $\ket{\psi_I}$ and connections $V_{IJ}$,
  which have truncated dimensions $\chi_I$ and $\chi_I \times \chi_J$, respectively,
  where typically $\chi_I \ll N$.
We use capital and lower-case Greek letters (e.g. $\ket{\Psi_I}$ vs $\ket{\psi_I}$)
  for wavefunctions in the full and truncated Hilbert spaces,
  respectively.
Similarly, we place hats on operators that act within the full Hilbert space (e.g. $\hat{U}_{IJ}$),
  while operators within the truncated Hilbert space (e.g. $V_{IJ}$) do not have hats.

In order to calculate expectation values of local operators $\hat{A}_I$ in the original Hilbert space,
  we must also define truncated operators, i.e. $\chi_I \times \chi_I$ matrices $A_I$, that act on the truncated Hilbert space.
Throughout this work, $\hat{A}_I$ always denotes an operator that acts within a patch $I$,
  and similar for $\hat{B}_J$, etc.
The truncated operators $A_I$ are notationally distinguished from the original operators $\hat{A}_I$ by the lack of a hat.
In \secref{sec:QGNFromTruncation}, we present a concrete mapping to obtain a QGN and truncated operators.
However, in many cases (e.g. \appref{app:KP} and \ref{app:Ising}) the truncated operators can be taken to be a simple Kronecker product,
  such as $\sigma_I^\mu = \mathbbl{1} \otimes \sigma^\mu$,
  where $\mathbbl{1}$ is an identity matrix and $\sigma^\mu$ is a $2\times 2$ Pauli matrix.

Ideally, we want the quantum gauge network to accurately encode approximate expectation values.
For example, if the QGN is an approximation for a wavefunction $\ket{\Psi}$,
  then we would like the QGN to accurately encode local expectation values; i.e. we want
  $\braket{\psi_I | A_I | \psi_I} \approx \braket{\Psi | \hat{A}_I | \Psi}$.
Similarly, we typically also want expectation values of string operators to also approximately match, e.g.
\begin{equation}
  \braket{\psi_I | A_I V_{IJ} B_J V_{JK} C_K | \psi_K} \approx
  \braket{\Psi | \hat{A}_I \hat{B}_J \hat{C}_K | \Psi} \label{eq:expValue}
\end{equation}
Note that in order to express a string operator that acts on multiple spatial patches using a QGN,
  it is essential to insert connections $V_{IJ}$ between operators and wavefunctions associated with different spatial patches.

\begin{figure}
  \centering
  \subfloat[\label{fig:snake1}]{\includegraphics{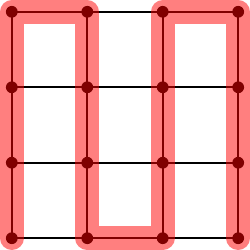}} \hspace{.15cm}
  \subfloat[\label{fig:snake2}]{\includegraphics{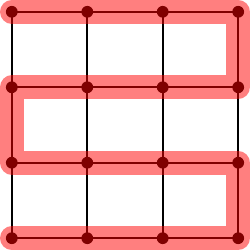}} \hspace{.15cm}
  \subfloat[\label{fig:snake3}]{\includegraphics{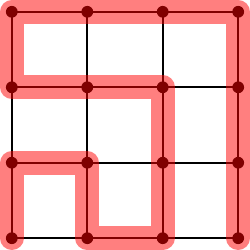}}
  \caption{%
    Three different paths that cover all lattice sites.
  }\label{fig:snake}
\end{figure}

Similar to the gauge picture of quantum dynamics,
  a density matrix can be extracted from a QGN.
Thus, a QGN most generally encodes a mixed state rather than a pure state.
For example, analogous to \eqnref{eq:gaugePictureRho} for a system of $n$ qubits,
  a density matrix can be approximately extracted from a QGN via
\begin{align}
  \hat{\rho} \approx 2^{-n} \sum_{\mu_1 \cdots \mu_n} &
    \hat{\sigma}_1^{\mu_1} \cdots \hat{\sigma}_n^{\mu_n} \label{eq:QGNRho}\\
    & \;\; \braket{\psi_1 | \, \sigma_1^{\mu_1} \, V_{1,2} \, \sigma_2^{\mu_2} \cdots \, V_{n-1,n} \, \sigma_n^{\mu_n} | \psi_n} \nonumber
\end{align}
  where $\hat{\sigma}_i^{\mu}$ are Pauli operators.
Here, we take each patch to consist of just a single qubit (for simplicity),
  and we use $I=1,\ldots,n$ to index the qubits/patches along a string of nearest-neighbors,
  e.g. as in \figref{fig:snake}.
However, due to the approximations induced by the QGN,
  different paths (e.g. those in \figref{fig:snake})
  can yield different density matrices.

Also similar to the gauge picture,
  quantum gauge networks exhibit a local gauge symmetry:
\begin{equation}
\begin{aligned}
  \ket{\psi_I} &\to \Lambda_I \ket{\Psi_I} \\
  V_{IJ} &\to \Lambda_I V_{IJ} \Lambda_J^\dagger \\
  A_I &\to \Lambda_I A_I \Lambda_I^\dagger
\end{aligned}
\end{equation}
  where $\Lambda_I$ is a unitary matrix.
Expectation values must be invariant under this symmetry.

Since local expectation values $\braket{\Psi | \hat{A}_I | \Psi} \approx \braket{\psi_I | A_I | \psi_I}$
  are encoded in the local wavefunctions $\ket{\psi_I}$,
  we can think of the local wavefunction as a purified reduced density matrix for a spatial patch.
The connections $V_{IJ}$ encode long-range correlations between the spatial patches.

It is desirable for a QGN to at least approximately obey the following consistency conditions
\begin{equation}
\begin{aligned}
  V_{IJ} \ket{\psi_J} &\approx \ket{\psi_I} \\
  V_{IJ} V_{JK} &\approx V_{IK}
\end{aligned} \label{eq:approxIdentities}
\end{equation}
Although it is easy to make the first relation exact,
  the second will typically only hold approximately.
Typically, a QGN will only possess a $V_{IJ}$ for nearby patches $I$ and $J$
  (and not for far away patches).
Thus, the second relation only applies if all three connections ($V_{IJ}$, $V_{JK}$, and $V_{IK}$) are contained in the QGN.
The connections $V_{IJ}$ should have singular values less than or equal to 1
  (to ensure that expectation values are never larger than the largest eigenvalue of the measured operator).

When $V_{IJ} \ket{\psi_J} = \ket{\psi_I}$ holds exactly,
  QGN connected correlation functions obey the usual identity (with a $V_{IJ}$ inserted):
\begin{equation}
\begin{aligned}
     &\bBraket{\psi_I \big| \big(A_I-\braket{A_I}\!\big) V_{IJ} \big(B_J-\braket{B_J}\!\big) \big| \psi_J} \\
  =\,&\braket{\psi_I | A_I V_{IJ} B_J | \psi_J} - \braket{A_I} \braket{B_J}
\end{aligned} \label{eq:QGNc}
\end{equation}
  where we abbreviate
  $\braket{A_I}=\braket{\psi_I|A_I|\psi_I}$ and
  $\braket{B_J}=\braket{\psi_J|B_J|\psi_J}$.
Therefore, if the QGN connected correlation function [\eqnref{eq:QGNc}] is small,
  we are guaranteed that 
  $\braket{\psi_I|A_I V_{IJ} B_J|\psi_J} \approx \braket{A_I} \braket{B_J}$,
  as one should expect.
The equations in this paragraph also hold for longer chains of connections;
  e.g. they also hold if we replace $V_{IJ}$ with $V_{IK} V_{KL} V_{LI}$.

\subsection{QGN from Truncation Maps}
\label{sec:QGNFromTruncation}

\begin{figure}
  \centering
  \includegraphics{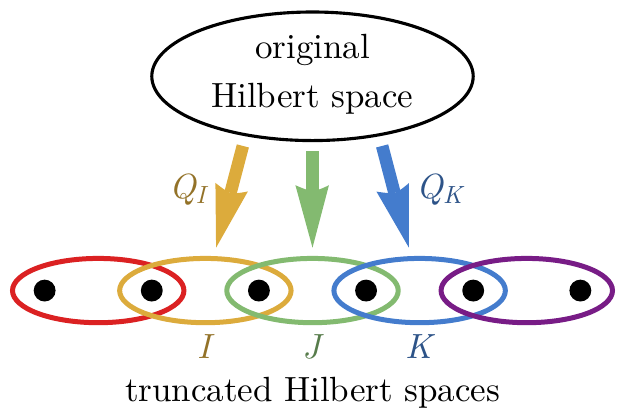}
  \caption{%
    Each truncation map $Q_I$ maps a subspace of the original Hilbert space on to the truncated Hilbert space associated with patch $I$.
  }\label{fig:truncationMaps}
\end{figure}

In this subsection, we study a concrete construction to obtain a quantum gauge network.
The input for this QGN construction is a wavefunction $\ket{\Psi}$, along with
  a \emph{truncation map} $Q_I$ for each patch of space.
If we want to construct a QGN from a density matrix instead,
  then $\ket{\Psi}$ should be chosen to be a purification of the density matrix.
The truncation maps are $\chi_I \times N$ matrices (where $N$ is the dimension of the full Hilbert space) that satisfy:
\begin{equation}
\begin{aligned}
  Q_I Q_I^\dagger &= \hat{\mathbbl{1}} \\
  Q_I^\dagger Q_I \ket{\Psi} &= \ket{\Psi}
\end{aligned} \label{eq:Q}
\end{equation}
Therefore $Q_I^\dagger$ is an isometry matrix whose image includes the wavefunction.
(A matrix $M$ is isometric if $M^\dagger M = \mathbbl{1}$.)
Intuitively, each $Q_I$ maps a select subspace of states into a truncated Hilbert space, as depicted in \figref{fig:truncationMaps}.
In the next subsection, we will explain how one can obtain useful truncation maps.

With this data, we can construct the following QGN:
\begin{equation}
\begin{aligned}
  \ket{\psi_I} &= Q_I \ket{\Psi} \\
  V_{IJ} &= Q_I Q_J^\dagger
\end{aligned} \label{eq:truncationMap}
\end{equation}
Note that $V_{II} = \mathbbl{1}$ due to \eqnref{eq:Q}.
Local operators $\hat{A}_I$ with support on a spatial patch $I$ are also truncated:
\begin{equation}
  A_I = Q_I \hat{A}_I Q_I^\dagger \label{eq:truncatedOp}
\end{equation}
Note that this equation can be used for both bosonic and fermionic operators.
In the limit that the bond dimension $\chi_I \to N$ approaches the full Hilbert space dimension $N$,
  this truncation mapping will result in a QGN that encodes correlation functions
  [e.g. \eqnref{eq:expValue}] exactly.

The operator norm of $V_{IJ}$ is bounded by $||V_{IJ}||_\text{op} \leq ||Q_I||_\text{op} ||Q_J^\dagger||_\text{op} = 1$.
Thus, the resulting connections $V_{IJ}$ have singular values that are less than or equal to 1.
One can verify that
\begin{equation}
  V_{IJ} \ket{\psi_J} = \ket{\psi_I} \label{eq:Vpsi}
\end{equation}
  from \eqnref{eq:approxIdentities} holds exactly.

In practice, the truncation mapping can only be directly applied for wavefunctions that are simple enough such that the truncation can be efficiently computed.
However, the construction is also useful for theoretically understanding how a QGN can encode a wavefunction.

A trivial example of a QGN can be obtained from a product state wavefunction
  $\ket{\Psi} = \ket{\psi_1} \otimes \ket{\psi_2} \otimes \cdots \ket{\psi_n}$.
The truncation maps can be chosen to be $Q_I = \ket{\psi_I} \bra{\Psi}$, with $I=1,\ldots,n$.
This results in a QGN with local wavefunctions $\ket{\psi_I}$
  and connections $V_{IJ} = \ket{\psi_I} \otimes \bra{\psi_J}$.
The truncated operators simply act within the on-site Hilbert space.
Additional examples of quantum gauge networks can be found in \appref{app:QGNExamples}.

Note that not all quantum gauge networks can be obtained from the truncation mapping.
For example, generating a QGN by sampling random numbers for $\ket{\psi_I}$ and $V_{IJ}$
  will result in a very unphysical QGN that is not consistent with any wavefunction.
This is in contrast to MPS or PEPS tensor networks,
  for which a random number initialization still returns a physical (although unnormalized) wavefunction.

In \appref{app:MPS}, we show how to obtain truncation maps from the canonical form of a matrix product state (MPS) \cite{OrusTNlong,ChanMPS,CiracMPS}.
If the MPS has bond dimension $\chi$,
  then the QGN will have bond dimensions equal to $d \chi^2$,
  where $d$ is the Hilbert space dimension at each site.
($d=2$ for qubits.)
The idea of the mapping is that the canonical form of an MPS can consist of a center tensor at $I$
  that is surrounded by isometric tensors.
The isometric tensors are then used to construct a truncation map $Q_I$,
  while the center tensor is a local wavefunction $\ket{\psi_I}$.

\subsection{Truncation Map Construction}
\label{sec:truncationMap}

The accuracy of the truncation critically depends on a good choice of truncation maps $Q_I$.
Suppose we want to choose truncation maps such that the quantum gauge network exactly encodes
  the expectation values of a chosen collection of operator strings.
For example, the expectation value of $\hat{A}_I \hat{B}_J \hat{C}_K$ is encoded exactly if
\begin{equation}
  \braket{\psi_I | A_I V_{IJ} B_J V_{JK} C_K | \psi_K} =
  \braket{\Psi | \hat{A}_I \hat{B}_J \hat{C}_K | \Psi} \label{eq:expExample}
\end{equation}
Below, we show that this exact encoding can be achieved by choosing truncation maps such that the image of each $Q_I^\dagger$
  is the span of certain strings of operators involved in the expectation values.

In general, we will find that the required bond dimension $\chi_I$ for each patch $I$ is bounded by $\chi_I \leq 1+2p_I$ [\eqnref{eq:chiBound}],
  where $p_I$ is the number of chosen operator strings (which we want to encode) that act on the patch $I$.
We will also find that bond dimension $O(M^k)$ [\eqnref{eq:OMk}] is sufficient to exactly encode all
  $2k$-point correlation functions of $M$ different operators.

\subsubsection{Warmup Example}

Before presenting a generic algorithm for obtaining the truncation maps, let us first discuss an instructive example.
Suppose we want to ensure that the QGN encodes the expectation value in \eqnref{eq:expExample} exactly
  for a particular choice of local operators ($\hat{A}_I$, $\hat{B}_J$, and $\hat{C}_K$)
  and spatial patches $I \neq J \neq K$.
Below, we show that any choice of truncation maps with the following images is sufficient:
\begin{equation}
\begin{aligned}
  \text{im}(Q_I^\dagger) &= \text{span}\big\{
    \ket{\Psi}, \;
    \hat{A}_I^\dagger \ket{\Psi} \big\} \\
  \text{im}(Q_J^\dagger) &= \text{span}\big\{
    \ket{\Psi}, \; \hat{A}_I^\dagger \ket{\Psi}, \;
    \hat{C}_K \ket{\Psi}
     \big\} \\
  \text{im}(Q_K^\dagger) &= \text{span}\big\{
    \ket{\Psi}, \;
    \hat{C}_K \ket{\Psi}  \big\}
\end{aligned} \label{eq:truncationExample}
\end{equation}
$\text{span}\big\{ \ket{\Psi_1},\ldots, \ket{\Psi_m} \big\}$ denotes the vector space spanned by the vectors $\ket{\Psi_1} ,\ldots, \ket{\Psi_m}$.
We set the bond dimensions to be equal to the vector space dimension of the images:
  $\chi_I = \text{dim}(\text{im}(Q_I^\dagger))$.
Specifying these images determines the truncation maps $Q_I$ up to a unitary gauge transformation $Q_I \to \Lambda_I Q_I$,
  where $\Lambda_I$ is a unitary matrix.
The choice of gauge does not affect QGN expectation values.
For examples of a quantum gauge networks obtained in this way,
  see \appref{app:QGNExamples}.

On a computer, a $Q_I^\dagger$ with the desired image can be calculated from the compact singular value decomposition $M_I = Q_I^\dagger S_I R_I$.
Here, $M_I$ is a matrix of column vector, which each encode one of the
  wavefunctions contained in the span of $\text{im}(Q_I^\dagger)$.
$S_I$ is a $\chi_I \times \chi_I$ diagonal matrix of nonzero singular values,
  and $R_I^\dagger$ is an isometry matrix.

Note that if $\ket{\Phi} \in \text{im}(Q_I^\dagger)$,
  then $Q_I^\dagger Q_I \ket{\Phi} = \ket{\Phi}$,
  which is a useful property.
This follows because $\ket{\Phi} \in \text{im}(Q_I^\dagger)$ implies that
  there exists $\ket{\phi}$ such that $Q_I^\dagger \ket{\phi} = \ket{\Phi}$, which then implies that
  $Q_I^\dagger Q_I \ket{\Phi} = Q_I^\dagger Q_I Q_I^\dagger \ket{\phi} = Q_I^\dagger \ket{\phi} = \ket{\Phi}$
  where the middle equality follows from $Q_I Q_I^\dagger = \hat{\mathbbl{1}}$ [\eqnref{eq:Q}].

For each patch, the image must contain $\ket{\Psi}$ so that \eqnref{eq:Q} is satisfied.
Next, we show that \eqnref{eq:truncationExample} is sufficient to exactly encode the expectation value in \eqnref{eq:expExample}:
\begin{align}
  & \braket{\psi_I | A_I V_{IJ} B_J V_{JK} C_K | \psi_K} \nonumber\\
 =& \braket{\Psi | Q_I^\dagger \,(Q_I \hat{A}_I Q_I^\dagger)\, V_{IJ} B_J V_{JK} \,(Q_K \hat{C}_K Q_K^\dagger)\, Q_K | \Psi} \nonumber\\
 =& \braket{\Psi | \hat{A}_I Q_I^\dagger V_{IJ} B_J V_{JK} Q_K \hat{C}_K | \Psi} \nonumber\\
 =& \braket{\Psi | \hat{A}_I Q_I^\dagger \,(Q_I Q_J^\dagger)\, B_J \,(Q_J Q_K^\dagger)\, Q_K \hat{C}_K | \Psi} \nonumber\\
 =& \braket{\Psi | \hat{A}_I Q_J^\dagger B_J Q_J \hat{C}_K | \Psi} \nonumber\\
 =& \braket{\Psi | \hat{A}_I Q_J^\dagger (Q_J \hat{B}_J Q_J^\dagger) Q_J \hat{C}_K | \Psi} \label{eq:expProof}\\
 =& \braket{\Psi | \hat{A}_I \hat{B}_J \hat{C}_K | \Psi} \nonumber
\end{align}
We used the identities $\ket{\psi_I} = Q_I \ket{\Psi}$ and $V_{IJ} = Q_I Q_J^\dagger$ [\eqnref{eq:truncationMap}],
  $A_I = Q_I \hat{A}_I Q_I^\dagger$ [\eqnref{eq:truncatedOp}],
  and $Q_I^\dagger Q_I \ket{\Phi} = \ket{\Phi}$ whenever $\ket{\Phi} \in \text{im}(Q_I^\dagger)$ in \eqnref{eq:truncationExample}.

Note that even if all the local operators commute,
  the path of the string matters.
For example, \eqnref{eq:truncationExample} does not guarantee that
  $\braket{\psi_J | B_J V_{JI} A_I V_{IK} C_K | \psi_K} =
  \braket{\Psi | \hat{B}_J \hat{A}_I \hat{C}_K | \Psi}$ even if $[\hat{A}_I, \hat{B}_J] = 0$.

If we want the QGN to encode expectation values for multiple operator strings,
  then we must calculate the images for each operator string,
  and then take the union.
For example, if two operator strings respectively require
  $\text{im}(Q_I^\dagger) \supseteq S_I^{(1)}$ and $\text{im}(Q_I^\dagger) \supseteq S_I^{(2)}$,
  then we need $\text{im}(Q_I^\dagger) \supseteq \text{span}(S_I^{(1)}, S_I^{(2)})$.

Equation~\eqref{eq:truncationExample} is not the unique choice for the images.
For instance, the images could be larger.
That is, it is sufficient to replace the equalities ``='' in \eqnref{eq:truncationExample} with superset relations ``$\supseteq$''.
Alternatively, we could replace $\text{im}(Q_J^\dagger)$ with
\begin{equation}
  \text{im}(Q_J^\dagger) = \text{span}\big\{
    \ket{\Psi}, \; \hat{B}_J \hat{C}_K  \ket{\Psi}, \;
    \hat{C}_K \ket{\Psi}
     \big\}
\end{equation}
And with this $\text{im}(Q_J^\dagger)$,
  we could then replace $\text{im}(Q_I^\dagger)$ with
\begin{equation}
  \text{im}(Q_I^\dagger) = \text{span}\big\{
    \ket{\Psi}, \;
    \hat{B}_J \hat{C}_K \ket{\Psi} \big\}
\end{equation}
The proof is similar to \eqnref{eq:expProof}.
By comparing these choices, we see that we have some freedom in choosing a midpoint in the string of operator products,
  which we make explicit in the generic case below.

\subsubsection{Generic Case}

More generally, suppose we want to ensure that a QGN exactly encodes
  the expectation value of $\hat{A}_{I_1}^{(1)} \hat{A}_{I_2}^{(2)} \cdots \hat{A}_{I_M}^{(M)}$,
  which is a product of $M$ local operators.
This expectation value is exactly encoded in the QGN if
\begin{equation}
\begin{aligned}
   & \bBraket{\psi_{I_1} \big| A_{I_1}^{(1)} V_{I_1,I_2} A_{I_2}^{(2)} V_{I_2,I_3} \cdots A_{I_M}^{(M)} \big| \psi_{I_M}} \\
  =& \bBraket{\Psi \big| \hat{A}_{I_1}^{(1)} \hat{A}_{I_2}^{(2)} \cdots \hat{A}_{I_M}^{(M)} \big| \Psi}
\end{aligned} \label{eq:exp1}
\end{equation}
  where $I_1,\ldots,I_m$ is a string of neighboring patches with $I_m \neq I_{m+1}$.
The exact encoding [\eqnref{eq:exp1}] is guaranteed by any choice of truncation maps with images that contain:
\begin{equation}
  \text{im}(Q_{I_m}^\dagger) \supseteq \begin{cases}
    \text{span}\big\{ \ket{\Psi}, \ket{\Psi_{m-1}^\text{L}}, \ket{\Psi_m^\text{L}} \big\} & m<m_0 \\
    \text{span}\big\{ \ket{\Psi}, \ket{\Psi_{m-1}^\text{L}}, \ket{\Psi_{m+1}^\text{R}} \big\} & m=m_0 \\
    \text{span}\big\{ \ket{\Psi}, \ket{\Psi_{m}^\text{R}}, \ket{\Psi_{m+1}^\text{R}} \big\} & m>m_0
  \end{cases}
  \label{eq:truncation1}
\end{equation}
  where we recursively define
\begin{equation}
\begin{aligned}
  \bKet{\Psi_0^\text{L}} &= \bKet{\Psi_{M+1}^\text{R}} = \bKet{\Psi} \\
  \bKet{\Psi_m^\text{L}} &= \hat{A}_{I_m}^{(m)\dagger} \bKet{\Psi_{m-1}^\text{L}} \\
  \bKet{\Psi_m^\text{R}} &= \hat{A}_{I_m}^{(m)} \bKet{\Psi_{m+1}^\text{R}}
\end{aligned}
\end{equation}
In \eqnref{eq:truncation1}, we are free to choose any half-integer $m_0 = 1,\tfrac{3}{2},\ldots,M$ between 1 and $M$.

As noted previously, if we want the QGN to encode expectation values for $M$ different operator strings,
  then for each image we obtain a list of conditions
  $\text{im}(Q_I^\dagger) \supseteq S_I^{(m)}$ from \eqnref{eq:truncation1} with $m=1,2,\ldots,M$.
We can then take the images to be the span of these vector spaces:
  $\text{im}(Q_I^\dagger) = \text{span}(S_I^{(1)}, S_I^{(2)}, \ldots S_I^{(M)})$.

We can easily bound the necessary bond dimensions $\chi_I$ needed to exactly encode the expectation values for many operator strings.
Let $p_I$ be the number of operator strings that involve the spatial patch $I$.
Each application of \eqnref{eq:truncation1} increases the dimension of $\text{im}(Q_I^\dagger)$ by at most 2
  [starting from $\chi_I=1$ since we always have $\ket{\Psi} \in \text{im}(Q_{I_m}^\dagger)$].
Therefore, this procedure results in bond dimensions of at most
\begin{equation}
  \chi_I \leq 1 + 2 p_I \label{eq:chiBound}
\end{equation}

To prove that \eqnref{eq:truncation1} implies \eqnref{eq:exp1},
  we first recall that $Q_I^\dagger Q_I \ket{\Phi} = \ket{\Phi}$ whenever $\ket{\Phi} \in \text{im}(Q_I^\dagger)$.
We thus obtain
\begin{equation}
\begin{aligned}
   & \bBra{\psi_{I_1}} A_{I_1}^{(1)} V_{I_1,I_2} \cdots A_{I_{\ceil{m_0}-1}}^{(\ceil{m_0}-1)} Q_{\ceil{m_0}-1} \\
  =& \bBra{\Psi_{I_1}} \hat{A}_{I_1}^{(1)} \cdots \hat{A}_{I_{\ceil{m_0}-1}}^{(\ceil{m_0}-1)}
\end{aligned} \label{eq:PsiLeft}
\end{equation}
and
\begin{equation}
\begin{aligned}
   & Q_{\floor{m_0}+1}^\dagger A_{I_{\floor{m_0}+1}}^{(\floor{m_0}+1)} \cdots V_{I_{M-1},I_M} A_{I_M}^{(M)} \bKet{\psi_{I_M}} \\
  =& \hat{A}_{I_{\floor{m_0}+1}}^{(\floor{m_0}+1)} \cdots \hat{A}_{I_M}^{(M)} \bKet{\Psi}
\end{aligned} \label{eq:PsiRight}
\end{equation}
by again using the identities $\ket{\psi_I} = Q_I \ket{\Psi}$, $V_{IJ} = Q_I Q_J^\dagger$, and $A_I = Q_I \hat{A}_I Q_I^\dagger$.
If $m_0$ is an integer, then inserting
  $V_{m_0-1,m_0} A_{m_0} V_{m_0,m_0+1} = Q_{m_0-1} Q_{m_0}^\dagger A_{m_0} Q_{m_0} Q_{m_0+1}^\dagger$
  and \eqsref{eq:PsiLeft} and \eqref{eq:PsiRight} into the first line of \eqnref{eq:exp1} yields
\begin{equation}
\begin{aligned}
   & \bBraket{\psi_{I_1} \big| A_{I_1}^{(1)} V_{I_1,I_2} \cdots A_{I_M}^{(M)} \big| \psi_{I_M}} \\
  =& \bBraket{\Psi \big| \hat{A}_{I_1}^{(1)} \cdots \hat{A}_{I_{m_0-1}}^{({m_0-1})} \big( Q_{m_0}^\dagger Q_{m_0} \hat{A}_{m_0} Q_{m_0}^\dagger Q_{m_0} \big)
     \\&\hspace{1cm} \hat{A}_{I_{m_0+1}}^{({m_0+1})} \cdots \hat{A}_{I_M}^{(M)} \big| \Psi} \\
  =& \bBraket{\Psi \big| \hat{A}_{I_1}^{(1)} \cdots \hat{A}_{I_M}^{(M)} \big| \Psi}
\end{aligned}
\end{equation}
If $m_0$ is a half-integer, then \eqnref{eq:exp1} follows immediately from
  $V_{\floor{m_0},\ceil{m_0}} = Q_{\floor{m_0}}^\dagger Q_{\ceil{m_0}}$ and \eqsref{eq:PsiLeft} and \eqref{eq:PsiRight}.
This completes the proof.

\subsubsection{Long-Range Correlation Functions}
\label{sec:correlators}

We can now show that quantum gauge networks can efficiently encode long-range correlation functions.
Suppose we want a QGN to exactly encode all two-point correlation functions
  for some arbitrary collection of operators $\hat{\tau}_i^\mu$ (indexed by $\mu$ and position $i$).
That is, suppose we want a QGN to satisfy
\begin{equation}
  \bBraket{\psi_I\big|\tau_{i\in I}^{\mu\dagger} V_{IJ}^\text{string} \tau_{j\in J}^\nu\big|\psi_J} =
  \bBraket{\Psi\big|\hat{\tau}_i^{\mu\dagger} \hat{\tau}_j^\nu\big|\Psi}
\end{equation}
Patches $I$ and $J$ can be any patches that respectively contain sites $i$ and $j$.
$\tau_{i\in I}^\mu = Q_I \hat{\tau}_i^\mu Q_I^\dagger$
  denotes a truncated operator on patch $I$.
Since patches $I$ and $J$ could be far apart,
  we insert a string of connections
\begin{equation}
  V_{IJ}^\text{string} = V_{IK_1} V_{K_1K_2} \cdots V_{K_l J}
\end{equation}
  to connect patches $I$ and $J$,
  where $(I,K_1,K_2, \ldots, K_l,J)$ is an arbitrary string of neighboring patches.
To exactly encode the above correlation functions,
  \eqnref{eq:truncation1} implies that
  it is sufficient for the images $\text{im}(Q_I^\dagger)$ to include the span of the actions of each operator on the wavefunction:
\begin{equation}
  \text{im}(Q_I^\dagger) \supseteq \text{span}\big\{\ket{\Psi}, \hat{\tau}_j^\mu\ket{\Psi} \text{ for each } \hat{\tau}_j^\mu \big\} \label{eq:im2point}
\end{equation}
If there are $M$ many operators $\hat{\tau}_i^\mu$,
  then the above image requires a QGN with bond dimension of at most $\chi_I = 1+M$.

More generally,
  suppose we want to exactly encode all $2k$-point correlation functions:
\begin{equation}
\begin{aligned}
  & \bBraket{\Psi\big| \hat{\tau}_{i_1}^{\mu_1\dagger} \cdots \hat{\tau}_{i_k}^{\mu_k\dagger}
                \hat{\tau}_{j_1}^{\nu_1} \cdots \hat{\tau}_{j_k}^{\nu_k} \big|\Psi} \\ &=
    \bBraket{\psi_{I_1}\big| \tau_{i_1\in I_1}^{\mu_1\dagger} V_{I_1,I_2}^\text{string} \cdots \tau_{i_k\in I_k}^{\mu_k\dagger}
    V_{I_k,J_1}^\text{string} \\ &\hspace{1.2cm}
                            \tau_{j_1 \in J_1}^{\nu_1} V_{J_1,J_2}^\text{string} \cdots \tau_{j_k\in J_k}^{\nu_k}\big|\psi_{J_k}}
\end{aligned}
\end{equation}
The left-hand-side is the expectation value of a product of $2k$ operators.
The right-hand-side can be any corresponding expectation value within a QGN for any valid strings of connections.
To exactly encode these correlation functions,
  \eqnref{eq:truncation1} implies that
  it is sufficient for the images to include the span of the actions of all products of up to $k$ operators:
\begin{equation}
  \text{im}(Q_I^\dagger) \supseteq \text{span}\left\{
    \text{all } \textstyle\prod_{r=1}^R \hat{\tau}_{j_r}^{\mu_r} \ket{\Psi} \text{ for } R=0,1,\ldots,k \right\} \label{eq:im k point}
\end{equation}
If there are $M$ many operators $\hat{\tau}_i^\mu$,
  then there are
\begin{equation}
  \chi = 1+M+M^2+\cdots+M^k \label{eq:OMk}
\end{equation}
  different operator products included in the span.
Therefore, a QGN with bond dimension of at most $\chi$ is sufficient to encode all $2k$-point correlation functions
  of $M$ many operators.

Although the bond dimension increases exponentially with $k$,
  typically only few-body operators can be measured in experiments.
Therefore, encoding $k$-point correlation functions with large $k$ may not be necessary.
On the other hand, for large systems, the number of operators $M$ is large,
  and it is advantageous that $\chi$ only increases as a polynomial for large $M$ and fixed $k$.

For a matrix product state (MPS),
  encoding all 2-point correlation functions generically requires an exponentially large bond dimension
  $\chi^\text{MPS} = 2^{M/6}$ for $M$ many operators on a chain of qubits.
For example, this exponential scaling is required when the wavefunction is a rainbow state.
However, a matrix product operator (MPO) with bond dimension $O(M)$
  is sufficient to encode all 2-point correlation functions.
See \appref{app:rainbow} for details.
Thus, QGN bond dimension scaling can be similar to that of an MPO.

See appendices~\ref{app:coherent} and \ref{app:slater} for analytical expressions of quantum gauge networks
  that exactly encode normal-ordered correlation functions of coherent boson wavefunctions and fermionic Slater determinant wavefunctions.

\section{Time Evolution Algorithm}
\label{sec:time}

Since the quantum gauge network is closely related to
  the gauge picture \cite{gaugePicture} of quantum dynamics,
  we can straight-forwardly modify the gauge picture
  to obtain an approximate algorithm for simulating quantum dynamics using a QGN.
To do this, we simply replace
  $\ket{\Psi_I} \to \ket{\psi_I}$,
  $\hat{U}_{IJ} \to V_{IJ}$, and
  $\hat{H}_{\langle I \rangle} \to H'_I$ (defined below)
  in the gauge picture equations of motion [\eqnref{eq:gaugePicture}] to obtain:
\begin{equation}
\begin{aligned}
  \partial_t \ket{\psi_I} &= -i H'_I \ket{\psi_I} \\
  \partial_t V_{IJ} &= - i H'_I V_{IJ} + i V_{IJ} H'_J
\end{aligned} \label{eq:EoM}
\end{equation}

If the Hamiltonian $\hat{H} = \sum_I \hat{H}_I$ [\eqnref{eq:H}]
  is a sum of local terms $\hat{H}_I$ each supported on a spatial patch $I$, then we define
\begin{equation}
  H'_I = \sum_J^{J \cap I \neq \emptyset} V_{IJ} \, H_J \, V_{JI} \label{eq:H'}
\end{equation}
$\sum_J^{J \cap I \neq \emptyset}$ sums over all patches $J$ that have nontrivial overlap with the spatial patch $I$.
$H'_I$ is analogous to $\hat{H}_{\langle I \rangle}$ in the gauge picture [\eqnref{eq:H'hat}].
Recall that $H_I$ is a $\chi_I \times \chi_I$ matrix in the truncated Hilbert space,
  while $\hat{H}_I$ is $N\times N$ and acts on the full $N$-dimensional Hilbert space.
$H_I$ can be obtained using the truncation mapping \eqref{eq:truncatedOp},
  which also holds when $\hat{H}_I$ is time-dependent.

For time-independent Hamiltonians, the QGN approximation
\begin{equation}
  E_\text{QGN}(t) = \sum_I \braket{\psi_I(t) | H_I | \psi_I(t)} \label{eq:EQGN}
\end{equation}
  of the energy expectation value $\braket{\Psi(t)|\hat{H}|\Psi(t)}$
  is conserved up to numerical integration errors for the time evolution in \eqnref{eq:EoM}.
See \appref{app:energyConservation} for a proof.

For a QGN with bond dimension $\chi$,
  the computation time of this algorithm is dominated by multiplication of $\chi \times \chi$ matrices.
Therefore, the computation time for each time step scales as $O(n_\text{V} \chi^3)$,
  where $n_\text{V}$ is the number of connections $V_{IJ}$ used by the QGN.
The memory cost scales as $O(n_\text{V} \chi^2)$ for storing the connections.\footnote{%
  Let $\widetilde{\chi}_{IJ}$ be the number of non-zero singular values of $V_{IJ}$.
  If $\widetilde{\chi}_{IJ} \ll \text{min}(\chi_I,\chi_J)$,
    then it is more efficient to decompose $V_{IJ} = V_I^{(J)} V_J^{(I)\dagger}$
    where $V_I^{(J)}$ is a $\chi_I \times \widetilde{\chi}_J$ isometric matrix.
  If all $\chi_I=\chi$ and $\widetilde{\chi}_{IJ}=\widetilde{\chi}$ are equal,
    and if the $H_I$ can be encoded as sparse matrices with $O(\chi)$ nonzero entries (which is typical),
    then the simulation and memory cost can respectively scale as $O(n_\text{V} \chi \widetilde{\chi}^2)$ and $O(n_\text{V} \chi \widetilde{\chi})$.
  This is asymptotically less costly when $\widetilde{\chi} \ll \chi$.
  However for the simulations in this work,
    this decomposition is not useful since we chose connections for which $\widetilde{\chi}/\chi$
    only ranges from about 0.75 (in one spatial dimension) to 0.5 (in three spatial dimensions).}
For local Hamiltonians, $n_\text{V}$ should be proportional to the number of lattice sites.

It is also possible to handle Hamiltonian terms that are not supported on a single spatial patch.
Consider the Hamiltonian
\begin{equation}
  \hat{H} = \sum_{J\cdots K} \sum_{\mu\cdots\nu}
      h_{J\cdots K}^{\mu\cdots\nu} \; \hat{\tau}^\mu_J \cdots \hat{\tau}^\nu_K \label{eq:Hgen}
\end{equation}
  which consists of a sum $\sum_{J\cdots K}$ over spatial patches $J \cdots K$.
$h_{J\cdots K}^{\mu\cdots\nu}$ are (generically time-dependent) real coefficients,
  and $\hat{\tau}^\mu_J$ denotes an operator indexed by $\mu$ with support on patch $J$.
For example, we could take the patches to consist of a single qubit,
  and the $\hat{\tau}^\mu_J$ could be Pauli operators.
For local Hamiltonians, $h_{J\cdots K}^{\mu\cdots\nu}$ will only be nonzero if $J\cdots K$ are close together.
Equation~\eqref{eq:H'} could be generalized to
\begin{align}
  H'_I =& \sum_{J\cdots K}^{(J\cup\cdots K)\cap I \neq \emptyset} \sum_{\mu\cdots\nu} \label{eq:H'gen}\\
        &\hspace{.8cm} \tfrac{1}{2} h_{J\cdots K}^{\mu\cdots\nu} \; (V_{IJ} \tau^\mu_J V_{JI}) \cdots (V_{IK} \tau^\nu_K V_{KI}) + h.c. \nonumber
\end{align}
  where $\sum_{J\cdots K}^{(J\cup\cdots K)\cap I \neq \emptyset}$ sums over spatial patches $J\cdots K$
  such that the union $J\cup\cdots K$ has nontrivial overlap with patch $I$.
``$h.c.$'' denotes the Hermitian conjugate of the preceding terms and ensures that $H'_I$ is Hermitian.
When $J\cdots K$ involves more than two patches,
  we must decide on an ordering of the $J\cdots K$ for each $I$.
\eqnref{eq:H'gen} follows from a similar generalized expression for $\hat{H}_{\langle I \rangle}$ in the gauge picture \cite{gaugePicture}
  after projecting onto the Hermitian part (via the $h.c.$)
  and replacing $\hat{U}_{IJ} \to V_{IJ}$ and $\hat{\tau}^\mu_J \to \tau^\mu_J$.
However, unlike for \eqnref{eq:H'}, the energy will not be conserved exactly for this choice of $H'_I$.

\subsection{Fermion Quench}
\label{sec:Fermi}

To benchmark this QGN algorithm, we simulate the dynamics of a quantum quench and compare to exact methods.
First, we study the quench dynamics for a model of spinless fermions in one, two, and three spatial dimensions.
In \appref{app:Ising},
  we also study the quench to a near-critical transverse field Ising model on a square lattice.
In all cases, we find that increasing the bond dimension increases the simulation accuracy.
This is as expected, since exact simulations of the gauge picture are reproduced
  once the bond dimension reaches the full Hilbert space dimension
  (or less when there are conserved quantities).

\begin{figure}
  \centering
  \includegraphics[width=6cm]{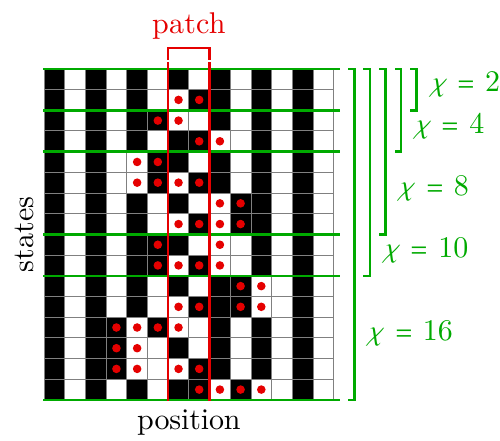}
  \caption{%
    States included in the truncated Hilbert space for a particular patch (labelled in red) of the 1D lattice.
    These are the states generated from five iterations of the algorithm described in the paragraph below \eqnref{eq:Fermi}.
    The initial state is shown in the top row,
      where white and black squares denote empty and filled fermions along the 1D chain
      (where we only show a subset of the chain near the patch).
    Below the initial state are additional states generated by the algorithm,
      where a red dot denotes a site with a different fermion number than the initial state.
    Each iteration ends on step (3),
      where we check if the number of states is sufficiently large.
    Each iteration is separated by a green line in the figure.
  }\label{fig:fermionStates}
\end{figure}

\begin{figure*}
  \centering
  \subfloat[22\label{fig:chainInit}]{\includegraphics{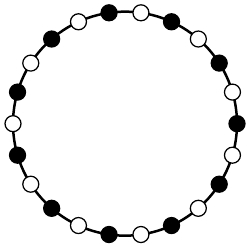}} \hspace{0.5cm}
  \subfloat[]{\includegraphics{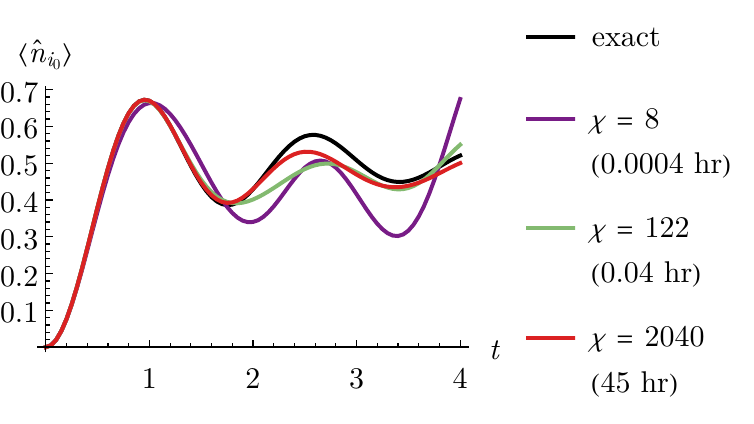}} \hspace{0.5cm}
  \subfloat[]{\includegraphics{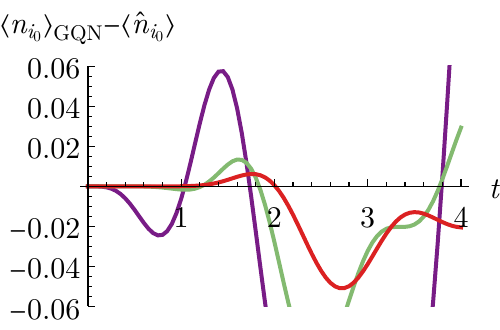}} \\
  \subfloat[$4\times 4$\label{fig:squareInit}]{\includegraphics{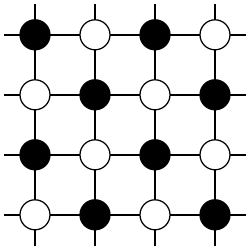}} \hspace{0.5cm}
  \subfloat[]{\includegraphics{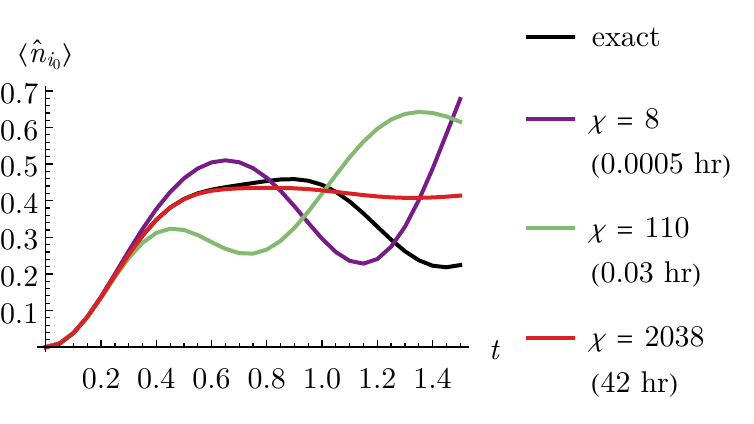}} \hspace{0.5cm}
  \subfloat[]{\includegraphics{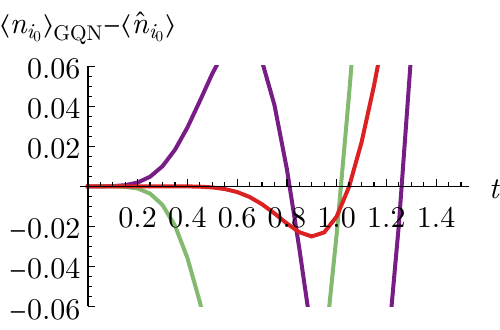}}
  \caption{%
    Simulation data for the time dynamics of the fermionic Hamiltonian \eqref{eq:Fermi} with $V=1$ following
      a quench from a checkerboard initial state in 1D and 2D periodic lattices.
    (a) An initial state of 11 spinless fermions (black disks) on a periodic chain of 22 sites (black and white disks).
    (b) Starting from this initial state, we show the fermion number expectation value $\braket{\hat{n}_{i_0}(t)}$ vs time $t$ at a site $i_0$
          with zero fermions at time $t=0$.
        Simulation results are shown for the exact value (black line) and
          the quantum gauge network (QGN) with different bond dimensions $\chi$ (colored lines).
        The legend also shows the number of CPU core hours used for each simulation.
    (c) The error of the QGN approximation $\braket{n_{i_0}}_\text{QGN}$ [\eqnref{eq:nQGN}] to the exact expression $\braket{\hat{n}_{i_0}}$.
    Panels (d-f) are similar, but for a periodic $4\times4$ square lattice.
    In both dimensions, we see that increasing the bond dimension increases the accuracy of the QGN simulations (for sufficiently small times).
  }\label{fig:FermiV1}
\end{figure*}

\begin{figure*}
  \centering
  \subfloat[22]{\includegraphics{fermionChain}} \hspace{0.5cm}
  \subfloat[]{\includegraphics{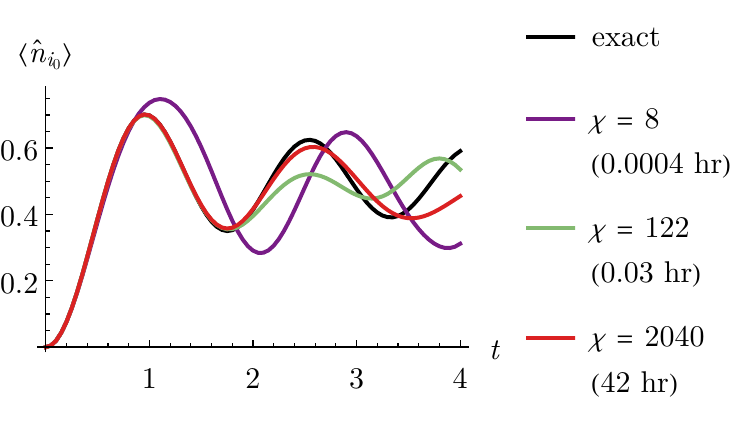}} \hspace{0.5cm}
  \subfloat[]{\includegraphics{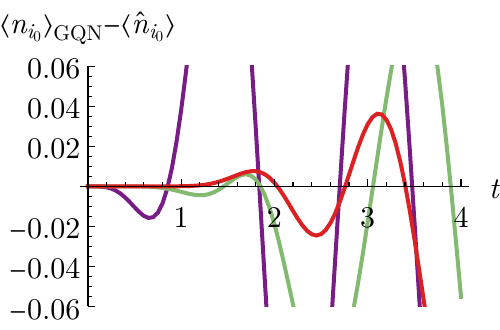}} \\
  \subfloat[$4\times4$]{\includegraphics{fermionSquare}} \hspace{0.5cm}
  \subfloat[]{\includegraphics{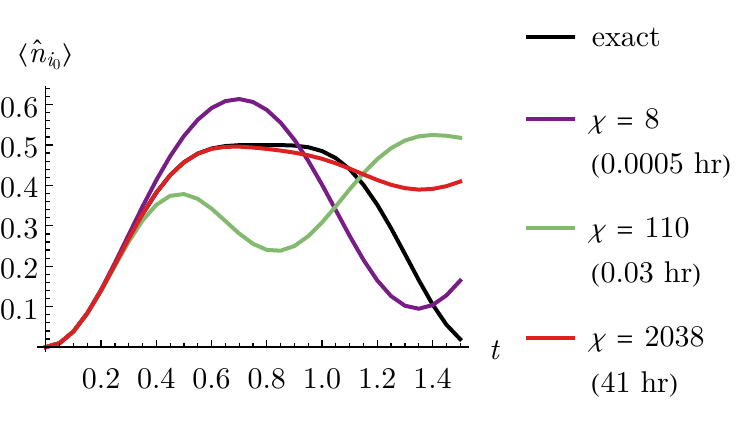}} \hspace{0.5cm}
  \subfloat[]{\includegraphics{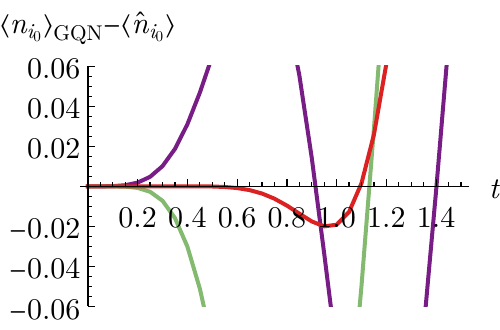}} \\
  \subfloat[$4\times 4\times 4$]{\includegraphics[width=70pt]{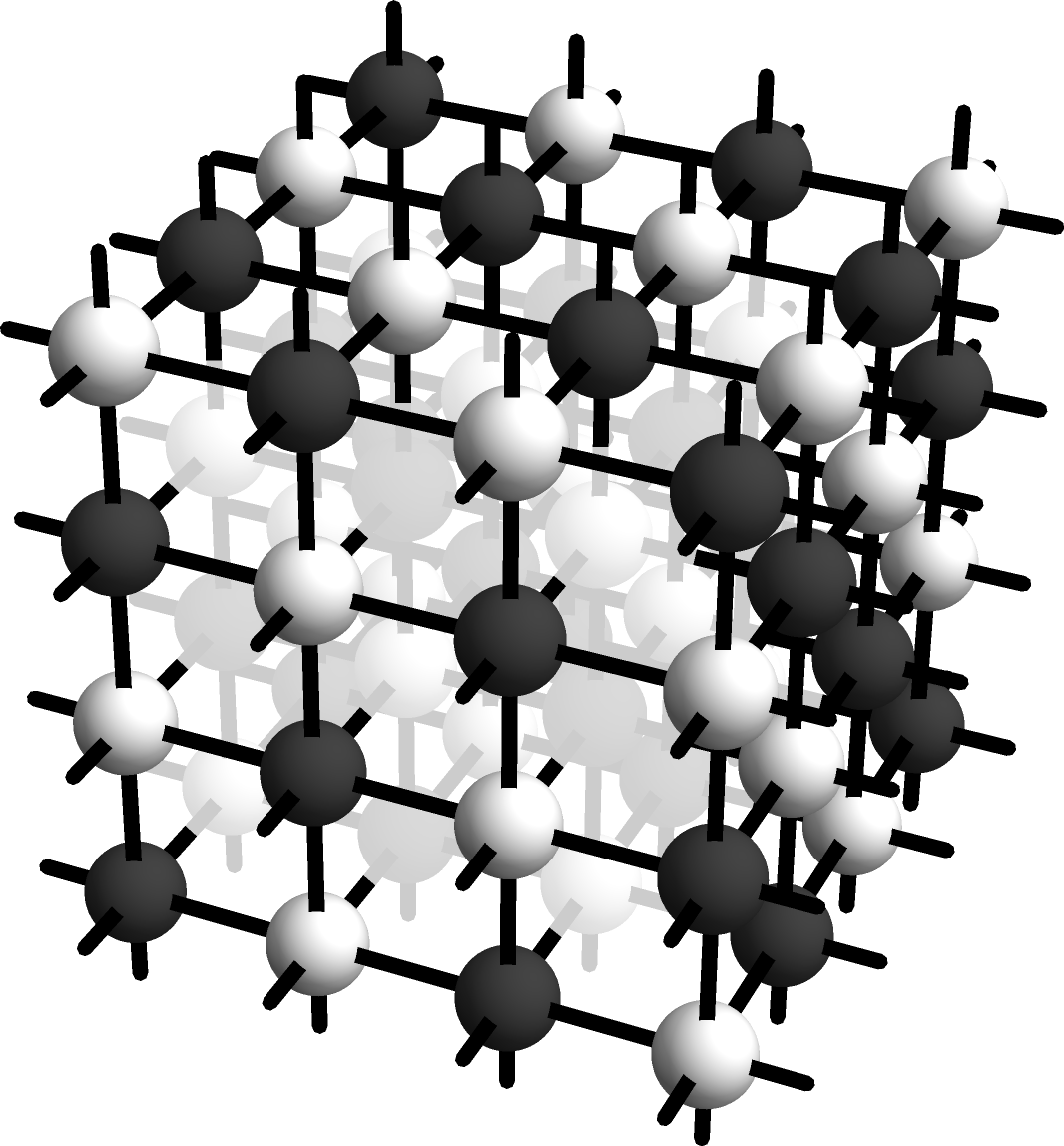}} \hspace{0.5cm}
  \subfloat[]{\includegraphics{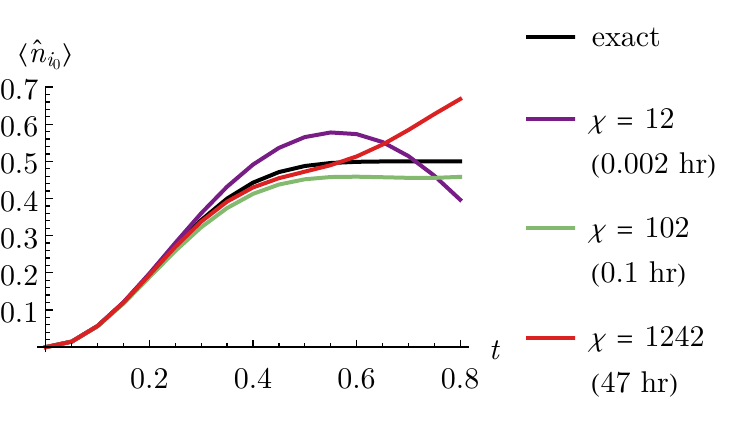}} \hspace{0.5cm}
  \subfloat[]{\includegraphics{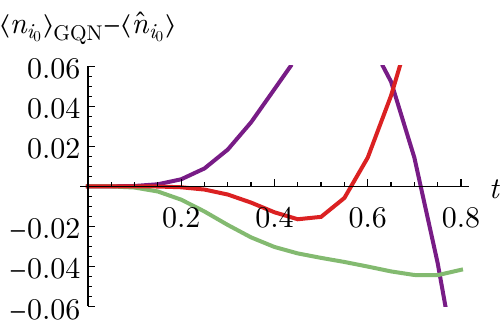}}
  \caption{%
    Panels (a-f) are the same as \figref{fig:FermiV1},
      except we simulate non-interacting fermions with $V=0$.
    By comparing to \figref{fig:FermiV1}, we see that the QGN simulates $V=0$ with roughly the same accuracy as $V=1$.
    Panels (g-i) are similar, but for a periodic $4\times4\times4$ cubic lattice.
  }\label{fig:FermiV0}
\end{figure*}

We initialize the system with a checkerboard pattern of spinless fermions,
  for which $\braket{\hat{n}_i(0)} = 1-\braket{\hat{n}_j(0)}$ at time $t=0$ for nearest-neighbor sites $i$ and $j$;
  as shown in \figsref{fig:chainInit} and \ref{fig:squareInit}.
We then time-evolve the system using the following Hamiltonian with nearest-neighbor hoppings and
  nearest-neighbor repulsive interactions:
\begin{equation}
\begin{aligned}
  \hat{H}^\text{Fermi} &= \sum_{\langle ij \rangle} \hat{H}^\text{Fermi}_{\langle ij \rangle} \\
  \hat{H}^\text{Fermi}_{I=\langle ij \rangle} &=
    - \hat{c}_i^\dagger \hat{c}_j - \hat{c}_j^\dagger \hat{c}_i + V \hat{n}_i \hat{n}_j
\end{aligned} \label{eq:Fermi}
\end{equation}
Each patch $I$ is composed of a pair of nearest-neighbor sites,
  which are summed over by $\sum_{\langle ij \rangle}$ in the first line.
$\hat{c}_i$ is a fermion annihilation operator,
  and $\hat{n}_i = \hat{c}_i^\dagger \hat{c}_i$ is the fermion number operator.

We initialize the QGN using truncation maps,
  as described in \secref{sec:QGNFromTruncation}.
Choosing the truncation maps $Q_I$ requires choosing a subspace of states to keep for each patch
  (i.e. a choice for the image of $Q_I^\dagger$, as in \secref{sec:truncationMap}).
We use the following method to select subspaces of states that we expect will acquire the highest weight after a short time evolution:
(1) We begin with images of $Q_I^\dagger$ that only contain the initial state.
(2) For each patch $I$ of nearest-neighbor sites,
      we add states to the image of $Q_I^\dagger$ that can be obtained from the current image by swapping the two sites within the patch.
(3) Stop if the bond dimension is sufficiently large.
(4) For each patch $I$, we add states that are included in patches $J$ that overlap with patch $I$.
(5) Go back to the second step.
See \figref{fig:fermionStates} for states that get included for the 1D lattice.

With this initialization algorithm, we do not choose the bond dimension $\chi$ precisely.
For example, \figref{fig:fermionStates} shows that only $\chi=2,4,8,10,16,\ldots$ are allowed for the 1D chain.
This restriction has the advantage that $\chi$ is chosen such that the QGN remains symmetric under lattice symmetries.
Each image consists of a span of certain eigenstates of the number operators $\hat{n}_i$.
Note that since the initial state has a definite fermion number,
  this procedure only adds states with the same fermion number,
  which is desirable since the Hamiltonian also conserves the fermion number.
If we were to continue the procedure until no additional states could be added,
  then we would add all states with the initial fermion number
  and the QGN simulation would be exact.

We compare the quantum gauge network results to the exact values.\footnote{%
  Exact expressions for small system sizes can be obtained by calculating the full wavefunction
    $\ket{\Psi(t)} = e^{-i \hat{H} t} \ket{\Psi(0)}$ using a sparse matrix representation for the Hamiltonian.}
Figure~\ref{fig:FermiV1} shows one-dimensional (1D) and two-dimensional (2D) simulation data with interaction strength $V=1$.
In both spatial dimensions, we find that the QGN can accurately simulate the dynamics for a short time,
  and the time over which the simulations are accurate increases with increasing bond dimension.
The most time-consuming simulation consumed 90 CPU hours,
  which took less than half a day on an 8-core laptop.

We calculate the number expectation value $\braket{\hat{n}_{i_0}(t)}$ for sites $i_0$ that initially have zero fermions.
In the QGN, these are estimated as follows:
\begin{equation}
  \braket{n_i}_\text{QGN} = \sum_{I \ni i}^\text{mean} \braket{\psi_I | n_{i \in I} | \psi_I} \label{eq:nQGN}
\end{equation}
$\sum_{I \ni i}^\text{mean}$ averages over all patches $I$ that contain the site $i$, and
  $n_{i \in I} = Q_I \hat{n}_i Q_I^\dagger$ is the truncated [\eqnref{eq:truncatedOp}] number operator at site $i$ for patch $I$.
In this example, $\braket{\psi_I | n_{i \in I} | \psi_I}$ is equal for all patches $I$ that contain site $i$ due to spatial symmetries.
However in other models with less symmetry, simulation errors can make these expectation values differ for different patches.

If we were to integrate the equations of motion exactly,
  then the energy expectation value [\eqnref{eq:EQGN}]
  would be conserved exactly.
Since exact integration is not practical,
  we use a modified RK4 Runge-Kutta method for numerical integration with time step $\delta_\text{t} = 0.05$.
Due to this approximation, the energy per site changed by at most $10^{-3}$ for all data shown.
See \appref{app:simulation} for more details.

In order to compare to exact methods in three dimensions (3D) with many sites,
  we repeat the comparison in \figref{fig:FermiV0} with no interactions ($V=0$).
Free (i.e. non-interacting $V=0$) fermion systems are efficient to simulate exactly \cite{Afoot:freeFermion} and are typically about as challenging for
  tensor network methods as interacting fermionic systems.
By comparing the 1D and 2D data in \figsref{fig:FermiV1} and \ref{fig:FermiV0},
  we indeed see that $V=1$ and $V=0$ appear to be roughly equally challenging for the QGN.
Therefore, we expect that the comparison we make for free fermions in 3D is representative of the interacting $V=1$ model
  (for which we could not perform exact simulations).
For short time evolutions, we find that the QGN simulation errors decrease as the bond dimension $\chi$ is increased.

For the free fermion simulations,
  the energy expectation value and total fermion number
  within the QGN appears to be conserved exactly
  (up to floating point precision).
We have not yet investigated why this conservation occurs in the free fermion system.
When interactions ($V\neq 0$) are included, the QGN conserves the energy expectation value up to numerical integration errors,
  and the QGN fails to conserve the total charge expectation value.

\section{Outlook}

Quantum gauge networks offer several advantages for simulating quantum dynamics,
  especially in comparison to many tensor network methods.
(1)~The computation time for simulating time dynamics scales as $\chi^3$, where $\chi$ is the bond dimension.
    Notably, this computation time does not increase with the spatial dimension,
      which makes quantum gauge networks a promising tool in two or more dimensions.
    Furthermore, the computational time per variational parameter is $\chi^3 / \chi^2 = \chi^{1.5}$,
      which is the same remarkably efficient ratio as MPS algorithms.
(2)~Fermionic models are simple to handle.
    (Unlike MERA or PEPS, fermionic swap gate \cite{fermionicTN,fermionicMERA} are not needed.)
(3)~The code is simple since the only tensors involved are vectors $\psi_I$ and matrices $V_{IJ}$,
      and the code does not get more complicated in larger spatial dimensions.
(4)~The energy expectation value can be conserved (up to integration error).
(5)~Lattice symmetries can be maintained exactly.
(6)~Time discretization errors are small in practice since the dynamics
    can be integrated using very accurate Runge-Kutta methods.
    (Trotter-Suzuki expansions \cite{TrotterTheory,ClusterTrotter} are not needed.)
(7)~Long-range interactions are easy to implement by simply including connections between distant patches.

There are numerous important future directions for the study of quantum gauge networks:
(1)~In comparison to other tensor networks, understanding quantum gauge networks is conceptually more demanding
      since the wavefunction is not directly encoded.
    The truncation mapping \eqref{eq:truncationMap} is an example for which we can understand how the QGN relates to a wavefunction.
    But we do not know how to tell if a given QGN is consistent with any choice of wavefunction and truncation maps.\footnote{%
      This problem may be related to the QMA complete \cite{MarginalQMA} quantum marginal problem \cite{QuantumMarginal,SymmetricQuantumMarginal,NRep}.}
    We also do not know to what extent the truncation mapping can produce all quantum gauge networks that are useful approximations for quantum wavefunctions.
(2)~Can we optimize a QGN to find approximate ground states or excited eigenstates?
    This is relatively challenging for quantum gauge networks because a QGN can encode unphysical states,
      which means that some sort of (possibly approximate) constraint must be imposed on the QGN during energy minimization.
(3)~Imaginary time evolution in the gauge picture does not yield ground state physics, as it does in Schr\"odinger's picture.
[The same is true for Heisenberg's picture where $\hat{A}^\text{H}(t=-i\tau) = e^{+\hat{H}\tau} \hat{A}(0) e^{-\hat{H}\tau}$, which isn't even Hermitian.]
Thus, imaginary time evolution may not appear to be a useful tool for obtaining approximate ground states using a QGN.
However, the imaginary time evolution of a Hamiltonian $H$ can equivalently be expressed as the real time evolution of the \emph{non-local} Hamiltonian $\hat{H}_\text{im}(t) = -i [\hat{H}, \ket{\Psi(t)}\bra{\Psi(t)}]$.
Although the non-locality is non-ideal, this kind of time evolution could be implemented in a QGN with all-to-all connectivity of the connections to yield a QGN ground state algorithm.
(4)~How well can a QGN encode topological states \cite{TopoZoo,GuTopoPEPS,VidalTopoPEPS,SETPEPS,TopoIsoTN}?

There are also many opportunities to significantly improve our QGN time evolution algorithm:
(5)~For the TEBD algorithm \cite{TEBD},
      it is straightforward to increase or optimally truncate the bond dimension during the simulation.
    In this work, we initialized the quantum gauge network using a simple basis of states in the number basis.
    We expect that this simple initialization is far from optimal, and that
      dynamically adding and removing more optimally chosen states throughout the time evolution could greatly improve simulation accuracy.
(6)~Can we obtain exact energy conservation when the Hamiltonian terms act on multiple spatial patches [as in \eqnref{eq:Hgen}]?
(7)~Can we obtain charge conservation for charge-conserving Hamiltonians?
(8)~The TEBD algorithm allows one to upper-bound the simulation error in terms of the truncation error.
    Can we also estimate the error of a QGN time evolution without comparing to other algorithms?
(9)~Could we simulate infinite system sizes when the Hamiltonian is translation-invariant.
(10)~Finally, after improving QGN algorithms,
      benchmarking QGN methods against other methods
      \cite{MPSDynamicsRev,TEBD,tDMRG,TDVPMPS,TDVPTherm,MPOTDVP,iPEPSDynamics,TreeTDVP,DMT,DAOE,KrylovTDVP,TreeDynamics,ThermalTN,isoTNDynamics,2DNNTime,MendlNNRealTime,PollmannNNScaling,timeDMET}
      will be useful.

We find it intriguing that the consistency conditions
  ($V_{IJ} \ket{\psi_J} \approx \ket{\psi_I}$ and  $V_{IJ} V_{JK} \approx V_{IK}$) in \eqnref{eq:approxIdentities}
  are precisely the equations for a classical lattice gauge theory \cite{LatticeGaugeTheory,latticeGaugeTheoryAndSpin} to be in its ground state when coupled to a Higgs field,
  where $V_{IJ}$ plays the role of the gauge connection and $\ket{\psi_I}$ is the Higgs field.
(See \appref{app:gauge} for more details.)
The primary difference is that in lattice gauge theory,
  the gauge connections are typically chosen to be unitary matrices.
But $V_{IJ}$ is not a unitary matrix;
  its singular values are only constrained to be less than or equal to 1.
With unitary gauge connections,
  no information is encoded locally in the classical ground state.
(Information is only encoded in non-contractible Wilson loops.)
It is remarkable that by simply relaxing the unitary constraint on the gauge connections (as in a QGN),
  grounds states of classical lattice gauge theory coupled to a Higgs field
  are capable of locally encoding approximate quantum wavefunctions.
Gauge theory plays a foundational role within the standard model of particle physics.
As such, it may be interesting to study the emergent physics of quantum gauge networks,
  viewed not as a computational tool,
  but instead as a new kind of classical lattice gauge theory
  that exhibits aspects of emergent quantum mechanics
  \cite{SlaglePreskill,AaronsonSureShor,Hooft,Adler,VanchurinEntropic,Nelson2012,InteractingWorlds}.

\begin{acknowledgments}
We thank Gunhee Park, Lesik Motrunich, Sayak Guha Roy, and Garnet Chan for helpful conversations.
K.S. was partially supported by
  the Walter Burke Institute for Theoretical Physics at Caltech; and
  the U.S. Department of Energy, Office of Science, National Quantum Information Science Research Centers, Quantum Science Center.
This research was supported in part by the National Science Foundation under Grants No. NSF PHY-1748958 and PHY-2309135; 
  the Gordon and Betty Moore Foundation Grant No. 2919.02;
  and the Welch Foundation Award No. C-2166.
\end{acknowledgments}


\bibliographystyle{quantum}
\bibliography{gaugeNetwork}

\begin{thebibliography}{10}
\bibitem{Afoot:freeFermion}
The time dynamics of a free fermion Hamiltonian $\hat{H} = \sum_{ij} h_{ij} \, \hat{c}_i^\dagger \hat{c}_j$ can be simulated exactly by calculating the time-evolved filled single-fermion wavefunctions $\ket{\phi_\alpha(t)} = e^{-iht} \ket{\phi_\alpha(0)}$. The wavefunction $\ket{\Psi} = \prod_\alpha^\text{filled} \big(\sum_i \braket{i|\phi_\alpha} \hat{c}_i^\dagger\big) \ket{0}$ is never explicitly calculated. $\prod_\alpha^\text{filled}$ denotes the product over the filled single-fermion wavefunctions, and $\ket{0}$ is the empty state with no fermions. Then $\braket{\hat{n}_i(t)} = \sum_\alpha^\text{filled} |\braket{i|\phi_\alpha(t)}|^2$, where $\ket{i}$ is the single-fermion wavefunction for a fermion at site $i$.
\end{thebibliography}

\appendix

\section{Higgsed Lattice Gauge Theory}
\label{app:gauge}

In this appendix, we review more details regarding the connection between Higgsed lattice gauge theory and quangum gauge networks.
Consider a lattice of vertices connected by edges, e.g. a triangular lattice.
Each vertex of the lattice hosts a Higgs field $\psi_i \in \mathbb{C}^N$ with $|\psi_i|=1$,
  i.e. a normalized complex vector with $N$ components.
Each edge of the lattice hosts a $\text{U}(N)$ gauge connection $U_{ij}$,
  i.e. an $N\times N$ unitary matrix.
The energy of a classical $\text{U}(N)$ lattice gauge theory coupled to a Higgs field (on a lattice with triangular plaquettes) is
\begin{equation}
  E = - \sum_{ijk}^\text{plaquettes} \tr U_{ij} U_{jk} U_{ki}
      - \sum_{\langle i j \rangle} \text{Re}(\psi_i^* \cdot U_{ij} \cdot \psi_j) \label{eq:latticeGauge}
\end{equation}
  where $U_{ji} = U_{ij}^\dagger$ and
  $\sum_{ijk}^\text{plaquettes}$ sums over all plaquettes of the lattice (in two or more dimensions).
$U_{ij} U_{jk} U_{ki}$ is a product of gauge fields around the edges of a plaquette
  (which we assumed to be a triangle only for notational simplicity).
$\sum_{\langle i j \rangle}$ sums over vertices $i$ and $j$ connected by an edge.
$\text{Re}(\psi_i^* \cdot U_{ij} \cdot \psi_j)$ denotes the real part of $\psi_i^* \cdot U_{ij} \cdot \psi_j$.

The energy of this classical lattice gauge theory is minimized when
  $U_{ij} U_{jk} = U_{ik}$ and $U_{ij} \cdot \psi_j = \psi_i$,
  which is analogous to the consistency conditions in \eqsref{eq:UIdentities} and \eqref{eq:approxIdentities}.
However, there is an important difference in each case:
(1) In the gauge picture,
  $\hat{U}_{IJ}$ is an $N\times N$ unitary matrix where $N$ is the full Hilbert space dimension,
  which increases exponentially with system size.
However in lattice gauge theory,
  $N$ is typically taken to be a fixed integer.
(2) In a quantum gauge network,
  $V_{IJ}$ is not a unitary matrix;
  its singular values are only constrained to be less than or equal to 1.

\section{Matrix Product State Mapping}
\label{app:MPS}

In this appendix, we show that
  any matrix product state (MPS) with bond dimension $\chi$
  can be mapped to a quantum gauge network with bond dimension $d \chi^2$,
  where $d$ is the Hilbert space dimension at each site.
($d=2$ for qubits).
Before explaining the mapping,
  we first briefly review MPS canonical forms.

\subsection{MPS Review}

A matrix product state is an efficient representation of a wavefunction,
  where the wavefunction amplitudes are given by matrix products. \cite{OrusTNlong,ChanMPS,CiracMPS}
A MPS is specified by a $\chi_i^\text{MPS} \times \chi_{i+1}^\text{MPS}$
  rectangular matrix $M_i^{(s_i)}$ for each site $i$ and local state $\ket{s_i}$.
Equivalently, if $s_i=1,2,\ldots,d_i$ can take on $d_i$ different states,
  then each $M_i$ can be viewed as a $\chi_i^\text{MPS} \times d_i \times \chi_{i+1}^\text{MPS}$ tensor.
We restrict $\chi_1^\text{MPS} = \chi_{n+1}^\text{MPS} = 1$.
The MPS wavefunction for a chain of $n$ sites is
\begin{align}
  \bKet{\Psi_\text{MPS}} &= \sum_{s_1 s_2 \cdots s_n} \tr\!\big( M_1^{(s_1)} M_2^{(s_2)} \cdots M_n^{(s_n)} \big) \bKet{s_1 s_2 \cdots s_n} \nonumber\\
  &\hspace{-.25cm} \begin{array}{c}\\{\scalebox{1.5}{=}}\\{\scriptstyle(n=5)}\end{array}
  \; \adjincludegraphics[valign=c]{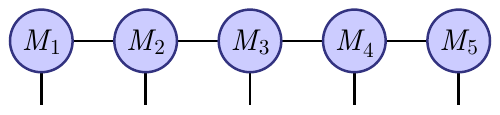} \label{eq:MPS}
\end{align}
The second line shows a tensor network diagram for $n=5$ sites.
The blue circles represent the tensors $M_i$;
  lines between tensors denote contracted indices;
  dangling lines denote uncontracted indices (the states $\ket{s_i}$ in this case);
  and we suppress the bond dimension $\chi_1^\text{MPS} = \chi_{n+1}^\text{MPS} = 1$ lines
  that are traced out.

There is a gauge redundancy $M_i^{(s_i)} \to \Lambda_i M_i^{(s_i)} \Lambda_{i+1}^\dagger$ (with unitary $\Lambda_i$)
  between neighboring matrices that does not affect the encoded wavefunction.
This redundancy is often used to compute a transformed MPS
  in a canonical form \cite{VidalTEBD,OrusTNlong} centered at a specific site $i$:
\begin{align}
  \bKet{\Psi_\text{MPS}} &= \sum_{s_1 \cdots s_n}
  \tr\!\Big( L_1^{(s_1)} \cdots L_{i-1}^{(s_{i-1})} C_i^{(s_i)} \\
  &\hspace{1.85cm} R_{i+1}^{(s_{i+1})} \cdots R_n^{(s_n)} \Big) \bKet{s_1 \cdots s_n} \nonumber\\
  &\hspace{-.25cm} \begin{array}{c}{\scriptstyle(i=3)}\\{\scalebox{1.5}{=}}\\{\scriptstyle(n=5)}\end{array}
   \; \adjincludegraphics[valign=c]{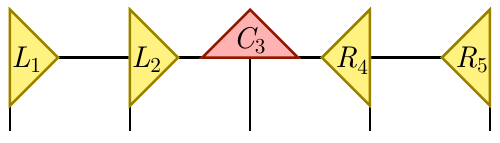} \nonumber
\end{align}
$L_i$, $C_i$, and $R_i$ are obtained from $M_i$
  using gauge transformations such that the following identities are obeyed:
\begin{equation}
\begin{aligned}
  \adjincludegraphics[valign=c]{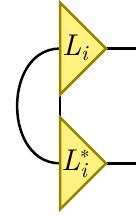} &= \sum_{s_i} L_i^{(s_i)\dagger} L_i^{(s_i)} = \mathbbl{1} \\
  \adjincludegraphics[valign=c]{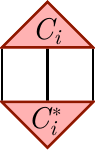} &= \sum_{s_i} \tr\!\big( C_i^{(s_i)\dagger} C_i^{(s_i)} \big) = 1 \\
  \adjincludegraphics[valign=c]{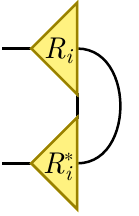} &= \sum_{s_i} R_i^{(s_i)} R_i^{(s_i)\dagger} = \mathbbl{1}
\end{aligned} \label{eq:canonicalConstraints}
\end{equation}
Thus, the $L_i$ and $R_i$ tensors are isometries,
  which pick out an orthonormal basis of states for the
  orthogonality center $C_i$, which is normalized like a wavefunction
  (but in a truncated Hilbert space).
One utility of the canonical form is that local expectation values are easy to compute:
\begin{equation}
  \braket{\Psi_\text{MPS} | \hat{A}_i | \Psi_\text{MPS}} =
    \sum_{s_is'_i} \tr\!\Big( C_i^{(s_i)\dagger} C_i^{(s'_i)} \Big) \braket{s_i|\hat{A}_i|s'_i}
\end{equation}
$\hat{A}_i$ is an operator that only acts on site $i$.


It is also possible to obtain multiple simultaneous canonical forms, one for each $i=1,2,\ldots,n$,
  while sharing the same isometries ($L_i$ and $R_i$).
These shared isometries obey
\begin{equation}
\begin{aligned}
  L_i^{(s_i)} C_{i+1}^{(s_{i+1})} &= C_i^{(s_i)} R_{i+1}^{(s_{i+1})} \\
  \adjincludegraphics[valign=c]{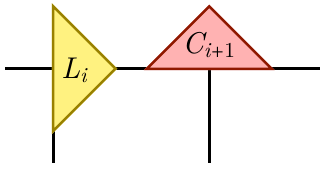} \;&=\; \adjincludegraphics[valign=c]{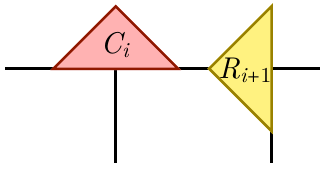}
\end{aligned} \label{eq:multicanonical}
\end{equation}
This simultaneous canonical form can be obtained by sweeping across the MPS multiple times using SVD decompositions
  or from the Vidal gauge \cite{VidalInfiniteMPS}.
The resulting bond dimensions of the $\chi_i^\text{MPS} \times d_i \times \chi_{i+1}^\text{MPS}$ tensors
  obey $\chi_i^\text{MPS} \leq d_i \chi_{i+1}^\text{MPS}$ and $\chi_i^\text{MPS} d_i \geq \chi_{i+1}^\text{MPS}$.

\subsection{Quantum Gauge Network from MPS}

We construct a quantum gauge network from an MPS using the truncation maps $Q_i$ defined in \secref{sec:truncationMap},
  which specify the truncated Hilbert space used by the QGN for each patch.
We choose $Q_i^\dagger$ to map the truncated Hilbert space
  of the MPS orthogonality center $C_i$ to the full Hilbert space:
\begin{align}
  Q_i^\dagger &= \sum_{s_1\cdots s_n}\sum_{\alpha,\beta}
    \Big( L_1^{(s_1)}\cdots L_{i-1}^{(s_{i-1})} \Big)_{1,\alpha} \\ &\hspace{2.2cm}
    \Big( R_{i+1}^{(s_{i+1})}\cdots R_n^{(s_n)} \Big)_{\beta,1} 
      \ket{s_1\cdots s_n}\bra{\alpha s_i\beta} \nonumber\\
  &\hspace{-.25cm} \begin{array}{c}{\scriptstyle(i=3)}\\{\scalebox{1.5}{=}}\\{\scriptstyle(n=5)}\end{array}
   \;\; \adjincludegraphics[valign=c]{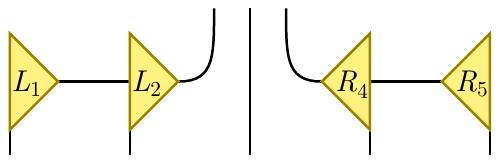} \nonumber
\end{align}
Thus, $Q_i^\dagger$ is just the canonical MPS centered at $i$,
  but with the orthogonality center $C_i$ removed.
Here, we choose the spatial patches to consists of just a single site,
  and we make no notational distinction between capital $(I,J,K)$ and lower case $(i,j,k)$ spatial indices letters.

Using \eqnref{eq:truncationMap} and the MPS canonical form identities [\eqnref{eq:canonicalConstraints}],
  we find that the local wavefunctions are equal to MPS orthogonality centers,
  while the connections are equal to tensor products of two MPS isometries:
\begin{align}
  \ket{\psi_i} &= \sum_{\alpha,s_i,\beta}
        \big( C_i^{(s_i)} \big)_{\alpha,\beta} \, \ket{\alpha s_i\beta} \nonumber\\
    &= \;\, \adjincludegraphics[valign=c]{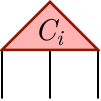} \nonumber\\
  V_{i,i+1} &= \sum_{\alpha,s_i,\alpha'} \sum_{\beta,s_{i+1},\beta'}
                 \big(L_i^{(s_i)}\big)_{\alpha\alpha'} \big(R_{i+1}^{(s_{i+1})}\big)^*_{\beta\beta'} \nonumber\\
            &\hspace{2.65cm} \ket{\alpha s_i \beta}\bra{\alpha' s_{i+1} \beta'} \\
    &= \quad \adjincludegraphics[valign=c]{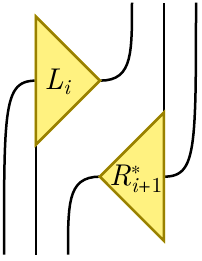} \nonumber
\end{align}
We use the simultaneous canonical forms,
  which obey \eqnref{eq:multicanonical} and guarantee that
  $V_{IJ} \ket{\psi_J} = \ket{\psi_I}$ [\eqnref{eq:Vpsi}]
  is satisfied.

The bond dimensions of the quantum gauge network are therefore
  $\chi_i = \chi_i^\text{MPS} d_i \chi_{i+1}^\text{MPS}$.
The singular values of the $\chi_i \times \chi_{i+1}$ matrix $V_{i,i+1}$ consist of
  $(\chi_{i+1}^\text{MPS})^2$ ones, while the rest are zero.
Thus, these $V_{i,i+1}$ are partial isometry matrices \cite{partialIsometry},
  which are matrices that obey $V V^\dagger V = V$
  (or equivalently matrices whose singular values are either zero or one).

\section{QGN Examples}
\label{app:QGNExamples}

In this appendix, we discuss several examples of quantum gauge networks.

\subsection{Mixed State Example}

As a simple example of a quantum gauge network,
  below we construct a QGN for the following mixed state of $n$ qubits:
\begin{equation}
\begin{aligned}
  \hat{\rho}^\text{mix}
    &= \tfrac{1}{2} \ket{\uparrow_1\uparrow_2\cdots\uparrow_n}
                    \bra{\uparrow_1\uparrow_2\cdots\uparrow_n} \\
  &\,+ \tfrac{1}{2} \ket{\downarrow_1\downarrow_2\cdots\downarrow_n}
                    \bra{\downarrow_1\downarrow_2\cdots\downarrow_n} \label{eq:rho}
\end{aligned}
\end{equation}

In order to apply the truncation mapping [\eqnref{eq:truncationMap}],
  we first find a purification of the density matrix:
\begin{equation}
\begin{aligned}
  \ket{\Psi^\text{mix}}
    &=   \tfrac{1}{\sqrt{2}} \ket{\uparrow_0\uparrow_1\uparrow_2\cdots\uparrow_n} \\
    &\,+ \tfrac{1}{\sqrt{2}} \ket{\downarrow_0\downarrow_1\downarrow_2\cdots\downarrow_n}
\end{aligned}
\end{equation}
The wavefunction $\ket{\Psi^\text{mix}}$ hosts an additional auxiliary qubit with index $I=0$.
In this example, we take the spatial patches to consist of a single qubit.
$\ket{\Psi^\text{mix}}$ is a purification of $\hat{\rho}^\text{mix}$ because tracing out the $I=0$ qubit
  yields the density matrix: $\hat{\rho}^\text{mix} = \tr_0 \ket{\Psi^\text{mix}}\bra{\Psi^\text{mix}}$.

We follow \secref{sec:truncationMap} to derive truncation maps $Q_I$
  by choosing the images $\text{im}(Q_I^\dagger)$
  to be carefully chosen subspaces of states.
To this end, we first note that any operator is a linear combination of Pauli strings,
  and the only Pauli strings with a nonzero expectation value for the state $\hat{\rho}^\text{mix}$
  are products of $\hat{\sigma}_I^\text{z} \hat{\sigma}_J^\text{z}$ operators.
We will therefore ensure that expectation values of
  products of $\hat{\sigma}_I^\text{z} \hat{\sigma}_J^\text{z}$ operators are retained by the truncation.
It turns out that this is sufficient to exactly encode all correlation functions for this example.
Equation~\eqref{eq:im2point} thus implies that we only need to include the action of a single $\hat{\sigma}_I^\text{z}$
  in the image of $Q_I^\dagger$:
\begin{equation}
\begin{aligned}
  \text{im}(Q_I^\dagger)
    &= \text{span}\big\{ \ket{\Psi^\text{mix}}, \; \hat{\sigma}_I^\text{z} \ket{\Psi^\text{mix}} \big\} \\
    &= \text{span}\big\{
       \ket{\uparrow_0\uparrow_1\cdots\uparrow_n}, \; \ket{\downarrow_0\downarrow_1\cdots\downarrow_n} \big\}
\end{aligned}
\end{equation}
This image has dimension $\chi_I=2$.
A natural gauge choice for $Q_I$ consistent with the above is:
\begin{equation}
\begin{aligned}
  Q_I &=   \ket{\uparrow}\bra{\uparrow_0\uparrow_1\uparrow_2\cdots\uparrow_n} \\
            &\,+ \ket{\downarrow}\bra{\downarrow_0\downarrow_1\downarrow_2\cdots\downarrow_n}
\end{aligned}
\end{equation}

Equation~\eqref{eq:truncationMap} then yields the following QGN:
\begin{equation}
\begin{aligned}
  \ket{\psi_I} &= \tfrac{1}{\sqrt{2}} \ket{\uparrow}
               + \tfrac{1}{\sqrt{2}} \ket{\downarrow} \\
  V_{IJ} &= \ket{\uparrow} \bra{\uparrow} + \ket{\downarrow} \bra{\downarrow}
\end{aligned}
\end{equation}
  with truncated operators [\eqnref{eq:truncatedOp}]
\begin{equation}
\begin{aligned}
  \sigma_I^\text{x} &= \sigma_I^\text{y} = 0 \\
  \sigma_I^\text{z} &= \ket{\uparrow}\bra{\uparrow} - \ket{\downarrow}\bra{\downarrow}
\end{aligned}
\end{equation}

This quantum gauge network exactly encodes all expectation values of the original reduced density matrix.
For example,
\begin{align}
  \braket{\psi_I | \sigma_I^\mu | \psi_I} &= \tr \hat{\rho}^\text{mix} \, \hat{\sigma}_I^\mu = 0 \\
  \braket{\psi_I | \sigma_I^\mu \, V_{IJ} \, \sigma_J^\nu | \psi_J}
     &= \tr \hat{\rho}^\text{mix} \, \hat{\sigma}_I^\mu \hat{\sigma}_J^\nu
      = \begin{cases} 1 & \mu = \nu = \text{z} \\
                      0 & \text{otherwise} \end{cases} \nonumber
\end{align}

\subsubsection{Kronecker Product Operators}
\label{app:KP}

If we want to preserve the algebra of more of the truncated operators,
  then we should include their action in the images.
For example, if we want to preserve the on-site algebra of the truncated Pauli operators,
  then we should instead choose:
\begin{equation}
\begin{aligned}
  \text{im}(Q_I^\dagger)
    &= \text{span}\big\{ \ket{\Psi^\text{mix}}, \; \hat{\sigma}_I^\text{x} \ket{\Psi^\text{mix}}, \; \hat{\sigma}_I^\text{y} \ket{\Psi^\text{mix}}, \; \hat{\sigma}_I^\text{z} \ket{\Psi^\text{mix}} \big\} \\
    &= \text{span}\big\{
       \ket{\uparrow_0\uparrow_1\cdots\uparrow_n}, \; \hat{\sigma}_I^\text{x} \ket{\uparrow_0\uparrow_1\cdots\uparrow_n}, \\ &\hspace{1.38cm}
       \ket{\downarrow_0\downarrow_1\cdots\downarrow_n}, \; \hat{\sigma}_I^\text{x} \ket{\downarrow_0\downarrow_1\cdots\downarrow_n} \big\}
\end{aligned}
\end{equation}
We can then pick the following truncation map:
\begin{equation}
\begin{aligned}
  Q_I &=   \ket{\uparrow_0\uparrow_I}\bra{\uparrow_0\uparrow_1\uparrow_2\cdots\uparrow_n} \\
            &\,+ \ket{\uparrow_0\downarrow_I}\bra{\uparrow_0\uparrow_1\uparrow_2\cdots\uparrow_n} \hat{\sigma}_I^\text{x} \\
            &\,+ \ket{\downarrow_0\uparrow_I}\bra{\downarrow_0\downarrow_1\downarrow_2\cdots\downarrow_n} \hat{\sigma}_I^\text{x} \\
            &\,+ \ket{\downarrow_0\downarrow_I}\bra{\downarrow_0\downarrow_1\downarrow_2\cdots\downarrow_n}
\end{aligned}
\end{equation}

The resulting QGN follows from \eqnref{eq:truncationMap}:
\begin{equation}
\begin{aligned}
  \ket{\psi_I} &= \tfrac{1}{\sqrt{2}} \ket{\uparrow_0\uparrow_I}
               + \tfrac{1}{\sqrt{2}} \ket{\downarrow_0\downarrow_I} \\
  V_{IJ} &= \ket{\uparrow_0\uparrow_I} \bra{\uparrow_0\uparrow_J} + \ket{\downarrow_0\downarrow_I} \bra{\downarrow_0\downarrow_J}
\end{aligned}
\end{equation}
Now the truncated Pauli operators are their natural Kronecker products:
\begin{equation}
\begin{aligned}
  \sigma_I^\text{x} &= \mathbbl{1}_0 \otimes \big( \quad\;\; \ket{\uparrow_I}\bra{\downarrow_I} + \;\; \ket{\uparrow_I}\bra{\downarrow_I} \big) \\
  \sigma_I^\text{y} &= \mathbbl{1}_0 \otimes \big( -i\ket{\uparrow_I}\bra{\downarrow_I} + i \ket{\uparrow_I}\bra{\downarrow_I} \big) \\
  \sigma_I^\text{z} &= \mathbbl{1}_0 \otimes \big( \quad\;\; \ket{\uparrow_I}\bra{\uparrow_I} + \;\; \ket{\downarrow_I}\bra{\downarrow_I} \big) \\
\end{aligned}
\end{equation}
  where $\mathbbl{1}_0 = \ket{\uparrow_0}\bra{\uparrow_0} + \ket{\downarrow_0}\bra{\downarrow_0}$.
These truncated operators obey their usual on-site algebra,
  e.g. $\sigma_I^\text{x} \sigma_I^\text{y} = i \sigma_I^\text{z}$.

\subsection{Cat State Example}
\label{sec:cat}

Encoding expectation values that act on many qubits in more than one spatial dimension can be less straight-forward.
For example, consider the following cat state
\begin{equation}
\begin{aligned}
  \ket{\Psi^\text{cat}}
    &=   \tfrac{1}{\sqrt{2}} \ket{\uparrow_1\uparrow_2\cdots\uparrow_n} \\
    &\,+ \tfrac{1}{\sqrt{2}} \ket{\downarrow_1\downarrow_2\cdots\downarrow_n}
\end{aligned}
\end{equation}
The only Pauli string expectation value that can distinguish $\ket{\Psi^\text{cat}}$ from
  the mixed state $\hat{\rho}^\text{mix}$ in \eqnref{eq:rho} is the highly-nonlocal product of Pauli $\hat{\sigma}^\text{x}$ operators on every qubit:
\begin{equation}
\begin{aligned}
  \tr \hat{\rho}^\text{mix} \prod_{I=1}^n \hat{\sigma}^\text{x}_I &= 0 \\
  \bigg\langle\Psi^\text{cat} \bigg|  \prod_{I=1}^n \hat{\sigma}^\text{x}_I \bigg| \Psi^\text{cat}\bigg\rangle &= 1
\end{aligned} \label{eq:catValue}
\end{equation}
Any Pauli string that does not act on all qubits will have an equal expectation value for $\hat{\rho}^\text{mix}$ and $\ket{\Psi^\text{cat}}$.

Now suppose that $\ket{\Psi^\text{cat}}$ is a wavefunction for a square lattice of qubits.
A quantum gauge network for $\ket{\Psi^\text{cat}}$ should reproduce the same nonlocal expectation value:
\begin{equation}
  \braket{\psi_1 | \sigma^\text{x}_1 V_{12} \sigma^\text{x}_2 \cdots V_{n-1,n} \sigma^\text{x}_n | \psi_n} = 1
\end{equation}
  where the sites $1, 2, \ldots, n$ snake across the square lattice,
  as depicted in \figref{fig:snake1}.
However, one may want other choices of paths [e.g. \figref{fig:snake2} or \ref{fig:snake3}]
  for this string operator to also lead to the same expectation value.
This can be achieved by adding these additional string operators to the procedure in \secref{sec:truncationMap}
  at the cost of increasing the bond dimension.
But if these additional string operators are not included in the QGN construction,
  then the expectation value of these excluded strings will not be encoded correctly.
This example demonstrates the issue that a quantum gauge network can seem to encode different values for the same nonlocal expectation value depending on the path chosen.

\subsection{Bosonic Coherent States}
\label{app:coherent}

The normal ordered expectation values of bosonic coherent states can be encoded within a quantum gauge network in a rather trivial way.
A bosonic coherent state is specified by complex numbers $\Theta_i$ and takes the following form:
\begin{equation}
  \ket{\Theta} = \exp\!\bigg(\!\sum_i \Theta_i \hat{b}_i^\dagger\bigg) \ket{0}
\end{equation}
$\hat{b}_i$ is a boson annihilation operator, which satisfies the commutation relations
  $[\hat{b}_i, \hat{b}_j] = 0$ and $[\hat{b}_i, \hat{b}_j^\dagger] = \delta_{ij}$,
  and $\ket{0}$ is the vacuum state with no bosons: $\hat{b}_i \ket{0} = 0$.

The coherent state is an eigenstate of the annihilation operators:
   $\hat{b}_i \ket{\Theta} = \Theta_i \ket{\Theta}$.
Therefore, if we only want the QGN to encode normal ordered expectation values,
  then \eqnref{eq:im k point} implies that the images of $Q_I^\dagger$ only need to contain one state: $\ket{\Theta}$.
We thus obtain a QGN with trivial bond dimensions $\chi_I=1$
  via the truncation map $Q_I = \ket{\widetilde{0}} \bra{\Theta}$.
Here, $\ket{\widetilde{0}}$ labels a state in a Hilbert space of dimension 1.
With this truncation mapping, the local wavefunctions are $\ket{\psi_I} = \ket{\widetilde{0}}$,
  and the connections are $V_{IJ} = \ket{\widetilde{0}}\bra{\widetilde{0}} = \mathbbl{1}$.
The truncated [\eqnref{eq:truncatedOp}] annihilation operator at a site $i$ in patch $I$ is simply
\begin{equation}
  b_{i \in I} = Q_I \hat{b}_i Q_I^\dagger = \Theta_i
\end{equation}
This QGN encodes all normal ordered expectation values exactly, e.g.
\begin{align}
  \bBraket{\Theta\big|\hat{b}_i^\dagger \hat{b}_j \big|\Theta}
    &= \bBraket{\psi_I \big| b_{i \in I}^\dagger V_{IJ} b_{j \in J} \big|\psi_J} \\
    &= \Theta_i^* \Theta_j \nonumber\\
  \bBraket{\Theta \big| \hat{b}_i^\dagger \hat{b}_j^\dagger \hat{b}_k \hat{b}_l \big| \Theta}
    &= \bBraket{\psi_I \big| b_{i \in I}^\dagger V_{IJ} b_{j \in J}^\dagger V_{JK} b_{k \in K} V_{KL} b_{l \in L} \big|\psi_L} \nonumber\\
    &= \Theta_i^* \Theta_j^* \Theta_k \Theta_l \nonumber
\end{align}

However, expectation values of operators that are not normal ordered are not encoded correctly by this QGN.
For example, $\braket{\psi_I | b_{i \in I} V_{IJ} b_{j \in J}^\dagger |\psi_J} = \Theta_i^* \Theta_j$
  while $\braket{\Theta | \hat{b}_i \hat{b}_j^\dagger | \Theta} = \delta_{ij} + \Theta_i^* \Theta_j$.
These additional expectation values could be encoded exactly by adding additional states to the images $\text{im}(Q_I^\dagger)$,
  as outlined in \secref{sec:correlators}.

\subsection{Fermion Slater Determinants}
\label{app:slater}

We can analytically construct a quantum gauge network that exactly encodes all
  normal-ordered two-fermion correlation functions $\braket{\hat{c}_i^\dagger\hat{c}_j}$
  for a fermionic Slater wavefunction.
If there are $n_\text{f}$ filled states,
  we can construct a QGN with bond dimension $\chi_I = 1+n_\text{f}$.
This is more efficient than the $1+M$ upper bound in \eqnref{eq:OMk},
  where $M=n$ is the number of operators whose correlation functions we wish to encode;
  here we consider all fermion annihilation operators $\hat{c}_i$ with $i=1,\ldots,n$.

A Slater determinant wavefunction can be expressed as
\begin{equation}
  \ket{\Phi} = \prod_{\alpha=1}^{n_\text{f}} \sum_{i=1}^n \Phi_{\alpha i} \, \hat{c}_i^\dagger \ket{0}
\end{equation}
  using second-quantized Fock states.
$i=1,\ldots,n$ indexes the $n$ different single-particle states.
$\hat{c}_i$ is a fermion annihilation operator, which satisfies the anticommutation relations
  $\{\hat{c}_i, \hat{c}_j\} = 0$ and $\{\hat{c}_i, \hat{c}_j^\dagger\} = \delta_{ij}$.
We fill $K$ many fermion orbitals, which are index by $\alpha$ and encoded by the matrix elements $\Phi_{\alpha i}$.
The orbitals are assumed to be orthonormalized: $\Phi \cdot \Phi^\dagger = \mathbbl{1}$.
The inner product of two Slater determinant wavefunctions is $\braket{\Phi|\Phi'} = \det(\Phi'\cdot\Phi^\dagger)$.

The action of an annihilation operator on a Slater determinant wavefunctions is
\begin{equation}
  \hat{c}_i \ket{\Phi} = \sum_{\alpha=1}^K (-1)^{\alpha-1} \Phi_{\alpha i} \ket{\Phi^-_\alpha} \label{eq:cPhi}
\end{equation}
We define
\begin{equation}
  \ket{\Phi^-_\alpha} = \prod_{\alpha' \neq \alpha} \sum_i \Phi_{\alpha'i} \hat{c}_i^\dagger \ket{0}
\end{equation}
  to be the Slater determinant wavefunction where we do not fill orbital $\alpha$,
  but the other $n_\text{f}-1$ orbitals are still filled.

Equation~\eqref{eq:im2point} implies that the images of $Q_I^\dagger$ only need to contain the states $\ket{\Phi}$ and $\ket{\Phi^-_\alpha}$.
Let $\ket{\phi}$ and $\ket{\phi^-_\alpha}$ label a basis of $1+n_\text{f}$ states for the local QGN Hilbert spaces.
Then we can choose truncation maps
\begin{equation}
  Q_I = \ket{\phi}\bra{\Phi} + \sum_{\alpha=1}^{n_\text{f}} \ket{\phi^-_\alpha} \bra{\Phi^-_\alpha}
\end{equation}
With this choice, the local wavefunctions are $\ket{\psi_I} = Q_I \ket{\Phi} = \ket{\phi}$,
  and the connections are $V_{IJ} = Q_I Q_J^\dagger = \mathbbl{1}$.
The truncated fermion operators follow from \eqnref{eq:cPhi}:
\begin{equation}
  c_{i\in I} = Q_I \hat{c}_i Q_I^\dagger = \sum_{\alpha=1}^K (-1)^{\alpha-1} \Phi_{\alpha i} \ket{\phi^-_\alpha} \bra{\phi}
\end{equation}
This QGN exactly encodes all normal-ordered two-fermion correlation functions:
\begin{equation}
\begin{aligned}
     \braket{\Phi | \hat{c}_i^\dagger\hat{c}_j | \Phi}
  &= \braket{\psi_I | c_{i\in I}^\dagger V_{IK_1} V_{K_1 K_2} \cdots V_{K_l J} c_{j\in J} | \psi_J} \\
  &= (\Phi^\dagger \cdot \Phi)_{ij}
\end{aligned}
\end{equation}
  where $V_{IK_1} V_{K_1 K_2} \cdots V_{K_l J}$ is any string of connections that connect patches $I$ and $J$.

Above, we only worked out analytical expressions for a QGN that exactly encodes two-fermion correlation functions.
But higher-point correlation functions are not encoded correctly.
For example, $\braket{\Phi | \hat{c}_i^\dagger \hat{c}_j^\dagger \hat{c}_k \hat{c}_l | \Phi}$ is not reproduced by the QNG because
  $\braket{\psi_I | c_{i\in I}^\dagger V_{IJ} c_{j\in J}^\dagger V_{JK} c_{k\in K} V_{KL} c_{l\in L} | \psi_L} = 0$
  since states with two fermions annihilated from $\ket{\Psi}$ are not included in the truncation.
However, analytical expressions for a QNG that encodes many-fermion correlation functions should also be possible.

The above QGN is rather trivial in the sense that the connections $V_{IJ}$ are all identity matrices.
This is because we did not take advantage of spatial locality.
If many of the fermion orbitals are spatially local,
  we expect that an approximate QGN encoding can be achieved with significantly smaller bond dimensions and non-identity $V_{IJ}$.

\subsection{Rainbow State}
\label{app:rainbow}

In \secref{sec:correlators}, we showed that all $2k$-point correlation functions of $M$ many operators
  can be encoded exactly by a QGN with bond dimension $O(M^k)$.
This is significantly more efficient than a  matrix product state (MPS),
  which can require bond dimension $\chi^\text{MPS} = 2^{n/2}$
  to encode all two-point correlation functions of certain states with $n$ qubits, e.g. the rainbow state.
In the $n$-qubit rainbow state, pairs of qubits $i$ and $n+1-i$ are maximally entangled in a Bell state.
The two-point correlation functions of the rainbow state are
   $\braket{\hat{\sigma}_i^\mu \hat{\sigma}_j^\nu} = - \delta_{\mu\nu} \delta_{n+1-i,j}$.
$\delta_{\mu\nu}$ denotes the Kronecker delta function.
The rainbow state is the unique state with these correlation functions.
Therefore, in order for an MPS to encode these 2-point correlation functions,
  the MPS must encode the rainbow state.
Encoding the rainbow state requires MPS bond dimension $\chi^\text{MPS} = 2^{n/2} = 2^{M/6}$,
  where $M=3n$ is the number of Pauli operators $\hat{\sigma}_i^\mu$ for $n$ qubits.

However, a matrix product operator (MPO) with bond dimension $\chi^\text{MPO} = 1+M/2$
  is sufficient to encode the 2-point correlation functions of the rainbow state by encoding the (unphysical) density matrix
  $\hat{\rho} = 2^{-n} \big[\hat{\mathbbl{1}} - \sum_{i=1}^{n/2} \sum_{\mu=\text{x},\text{y},\text{z}} \hat{\sigma}_i^\mu \hat{\sigma}_{n+1-i}^\mu\big]$.
This density matrix is unphysical because it has negative eigenvalues.
(A QGN bond dimension of $\chi = 1+M/2$ would also be sufficient for this example if we restrict the allowed operator strings to never change direction.)

\section{Energy Conservation}
\label{app:energyConservation}

Below, we prove that the QGN equations of motion [\eqnref{eq:EoM}] preserve the energy expectation value [\eqnref{eq:EQGN}] exactly
  when the local Hamiltonian terms $\hat{H}_I$ are time-independent and each supported on a single spatial patch
  [as in \eqsref{eq:H} and \eqref{eq:H'}].
\begin{align}
   & \partial_t \sum_I \braket{\psi_I | H_I | \psi_I} \nonumber \\
  =& \sum_I i \braket{\psi_I | [H'_I, H_I] | \psi_I} \nonumber \\
  =& \sum_{IJ}^{I\cap J \neq \emptyset} i \braket{\psi_I | [V_{IJ} H_J V_{JI}, H_I] | \psi_I} \label{eq:conservedEnergy} \\
  =& \sum_{IJ}^{I\cap J \neq \emptyset}
     \frac{i}{2} \braket{\psi_I | [V_{IJ} H_J V_{JI}, H_I] | \psi_I} \nonumber \\ &\quad\;\; +
     \frac{i}{2} \braket{\psi_J | [V_{JI} H_I V_{IJ}, H_J] | \psi_J} \nonumber \\
  =& \sum_{IJ}^{I\cap J \neq \emptyset}
     \frac{i}{2} \braket{\psi_I | [V_{IJ} H_J V_{JI}, H_I] + [H_I, V_{IJ} H_J V_{JI}] | \psi_I} \nonumber \\
  =& \, 0 \nonumber
\end{align}
The first three equalities respectively follow from \eqnref{eq:EoM} for $\partial_t \ket{\psi_I}$;
  \eqnref{eq:H'} for $H'_I$;
  and symmetrizing the sum over $I \leftrightarrow J$.
$\sum_{IJ}^{I\cap J \neq \emptyset}$ denotes the sum over all patches $I$ and $J$ that have nonzero overlap.
The final equality follows from the antisymmetry of the commutator.
The second to last equality follows from:
\begin{equation}
\begin{aligned}
 & \braket{\psi_J | [V_{JI} H_I V_{IJ}, H_J] | \psi_J} \\
=& \braket{\psi_J | V_{JI} H_I V_{IJ} H_J - H_J V_{JI} H_I V_{IJ} | \psi_J} \\
=& \braket{\psi_I | H_I V_{IJ} H_J V_{JI} - V_{IJ} H_J V_{JI} H_I | \psi_I} \\
=& \braket{\psi_I | [H_I, V_{IJ} H_J V_{JI}] | \psi_I}
\end{aligned}
\end{equation}
  which follows from $V_{IJ} \ket{\psi_J} = \ket{\psi_I}$ [\eqnref{eq:Vpsi}].

\section{Ising Model Quench}
\label{app:Ising}

In this appendix, we benchmark the quantum gauge network by studying the dynamics following
  a quench to a near-critical Ising model.
We start form the initial state
  $\ket{\Psi(0)} = \otimes_i \ket{\rightarrow_i}$ where all $\braket{\sigma_i^\text{x}} = 1$,
  and then we time evolve with a near-critical transverse field Ising Hamiltonian
\begin{equation}
  \hat{H}^\text{Ising} = - \sum_{\langle ij \rangle} \hat{\sigma}_i^\text{z} \hat{\sigma}_j^\text{z} - h \sum_i \hat{\sigma}_i^\text{x} \label{eq:Ising}
\end{equation}
  with $h=3$ on a two-dimensional $4\times4$ square lattice with periodic boundary conditions.
(The critical point is at $h_c\approx3.045$ \cite{Ising2D}.)
This system size is chosen so that we can compare to exact methods
  that calculate the full wavefunction $\ket{\Psi(t)} = e^{-i \hat{H}^\text{Ising} t}\ket{\Psi(0)}$.

In order to make use of \eqsref{eq:H} and \eqref{eq:H'},
  we define the Hamiltonian $\hat{H}_I$ on each spatial patch to be
\begin{equation}
  \hat{H}_{I=\langle ij \rangle}^\text{Ising} = - \hat{\sigma}_i^\text{z} \hat{\sigma}_j^\text{z} - \tfrac{1}{4} h \, (\hat{\sigma}_i^\text{x} + \hat{\sigma}_j^\text{x})
\end{equation}
We take each spatial patch $I$ to be a pair of nearest-neighbor sites $\langle ij \rangle$.
Note that in the sum $\hat{H} = \sum_I \hat{H}_I$ from \eqnref{eq:H},
  each site is summed over four times on a square lattice;
  thus we require the above $\tfrac{1}{4}$ factor in front of $h$.

We initialize the QGN using truncation maps
  (as described in \secref{sec:QGNFromTruncation}), which are chosen
  using a method similar to the one described in \secref{sec:Fermi}.
However in this spin model, we do not have a conserved charge.
Therefore, we modify step 2 of the method in \secref{sec:QGNFromTruncation} [paragraph below \eqnref{eq:Fermi}] to the following:
(2) For each patch $I$,
      we add states to the image of $Q_I^\dagger$ that can be obtained from the current image of $Q_I^\dagger$ by acting with Pauli operators within the patch $I$.

The truncation at each patch $I$ only retains states
  consisting of a span of eigenstates of the $\hat{\sigma}_i^\text{x}$ operators.
With a natural gauge choice for the truncation maps,
  the truncated Pauli operators take the form of a Kronecker product:
\begin{equation}
  \sigma_{i\in I}^\mu = Q_I \hat{\sigma}_i^\mu Q_I^\dagger = \mathbbl{1} \otimes \sigma^\mu
\end{equation}
$\sigma_{i\in I}^\mu$ is the truncated [\eqnref{eq:truncatedOp}] Pauli operator at site $i$ for patch $I$,
  and $\sigma^\mu$ is a $2\times 2$ Pauli matrix.

\begin{figure}
  \centering
  \subfloat[]{\includegraphics{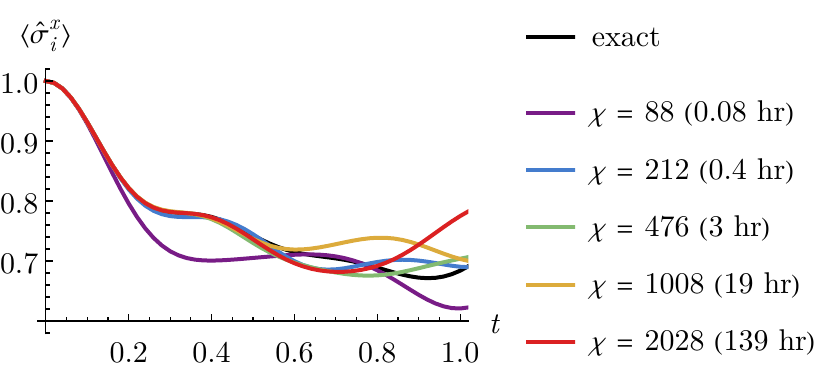}} \\
  \subfloat[]{\includegraphics{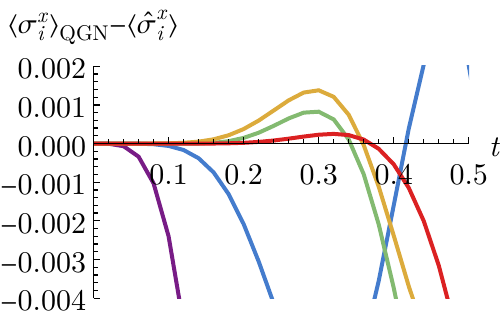}} \hspace{1cm}
  \caption{%
    Simulation data for the time dynamics of the near-critical transverse field Ising Hamiltonian \eqref{eq:Ising} on a periodic $4\times 4$ square lattice
      following a quench from the state $\ket{\Psi(0)} = \otimes_i \ket{\rightarrow_i}$.
    (a) The quantum gauge network (QGN) approximation for $\braket{\sigma_i^\text{x}}$ vs time $t$
        for different bond dimensions $\chi$ (colored lines) vs the exact value (black line).
    The legend also shows the number of CPU core hours used for each simulation.
    Due to symmetry, the $\braket{\sigma_i^\text{y}}$ and $\braket{\sigma_i^\text{z}}$ expectation values (not shown) are exactly zero for all time
      for both the QGN and the exact value.
    (b) The error $\braket{\sigma_i^\text{x}}_\text{QGN} - \braket{\sigma_i^\text{x}}_\text{exact}$
        of the QGN approximation to $\braket{\sigma_i^\text{x}}$.
    The error stays small for longer times as we increase the bond dimension $\chi$.
  }\label{fig:Ising}
\end{figure}

In \figref{fig:Ising}, we show QGN simulation data for Pauli expectation values $\braket{\hat{\sigma}^\mu_i(t)}$ and compare to the exact values.
We see that the simulation errors expectation value decrease as we increase the bond dimension.

In the QGN, the expectation values are estimated as
\begin{equation}
  \braket{\hat{\sigma}_i^\mu}_\text{QGN} = \sum_{I \ni i}^\text{mean} \braket{\psi_I | \sigma_{i \in I}^\mu | \psi_I}
\end{equation}
  where $\sum_{I \ni i}^\text{mean}$ averages over all patches $I$ that contain the site $i$.
In this example, $\braket{\psi_I | \sigma_{i\in I}^\mu | \psi_I}$ is equal for all patches $I$ that contain site $i$ due to spatial symmetries.
However in other models with less symmetry, simulation errors can make these expectation values differ for different patches.

If we were to integrate the equations of motion exactly,
  then the energy expectation value [\eqnref{eq:EQGN}]
  would be conserved exactly.
Since exact integration is not practical,
  we use a modified RK4 Runge-Kutta method for integration with time step $\delta_\text{t} = 0.02$.
Due to this approximation, the energy per site changed by $5\times10^{-4}$ and $1\times10^{-4}$
   for the $\chi=88$ and $2028$ simulations, respectively.
See \appref{app:simulation} for more details.

\section{Modified Runge-Kutta}
\label{app:simulation}

We use a modified RK4 Runge-Kutta method to integrate the differential equations.
RK4 is a forth-order Runge-Kutta method that results in an $O(\delta_\text{t}^4)$ error at time $t\sim1$,
  where $\delta_\text{t}$ is the time step size.
However, the straight-forward application of Runge-Kutta will not preserve $V_{IJ} \ket{\psi_J} = \ket{\psi_I}$ [\eqnref{eq:Vpsi}] exactly;
  it will only be preserved up to $O(\delta_\text{t}^4)$ error.
In this work, we chose to modify the Runge-Kutta method slightly such that $V_{IJ} \ket{\psi_J} = \ket{\psi_I}$ is preserved exactly
  (i.e. up to floating point precision).
We end up making an additional approximation that increases the simulation error to $O(\delta_\text{t}^3)$
  (which we were satisfied with).
It would be useful to improve the approximation such that $O(\delta_\text{t}^4)$ and smaller errors can be achieved
  while maintaining $V_{IJ} \ket{\psi_J} = \ket{\psi_I}$.

Instead of integrating $\partial_t \ket{\psi_I}$ and $\partial_t V_{IJ}$ directly at each time step,
  we use a modified Runge-Kutta method to obtain estimates for the unitary evolution
\begin{equation}
  U_I^\text{RK}(t+\delta_\text{t},t) = e^{-i \delta_\text{t} G_I^\text{RK}(t)}
\end{equation}
We then update the QGN from time $t$ to $t+\delta_\text{t}$ as follows:
\begin{equation}
\begin{aligned}
  \ket{\psi_I(t+\delta_\text{t})} &= U_I^\text{RK}(t+\delta_\text{t},t) \ket{\psi_I(t)} \\
  V_{IJ}(t+\delta_\text{t}) &= U_I^\text{RK}(t+\delta_\text{t},t) V_{IJ}(t) U_J^\text{RK}(t+\delta_\text{t},t)^\dagger
\end{aligned} \label{eq:QGNRK}
\end{equation}
We obtain
\begin{equation}
  G_I^\text{RK}(t) = \sum_{k=1}^s b_k \widetilde{G}_I^{(k)}(t)
\end{equation}
  using the Runge-Kutta coefficients $b_k$, where
  $s$ is the number of Runge-Kutta stages.
$s=4$ for RK4.

In order to calculate $\widetilde{G}_I^{(k)}(t)$,
  we first recursively define $G_I^{(k)}(t_k)$ (without the tilde) as\footnote{%
    Equivalently, the right-hand-side of \eqnref{eq:GIk} is $H'_I$ from \eqnref{eq:H'} at time $t_k$
      evaluated using the QGN that is updated from time $t$ to $t_k$ by $\widetilde{U}^{(k)}_I(t_k,t)$.}
\begin{equation}
  G_I^{(k)}(t_k) =
    \sum_J^{J \cap I \neq \emptyset} V_{IJ}^{(k)}(t_k) \, H_J(t_k) \, V_{JI}^{(k)}(t_k)
    \label{eq:GIk}
\end{equation}
  where $t_k = t+c_k \delta_\text{t}$ and
\begin{align}
  V_{IJ}^{(k)}(t_k) &= \widetilde{U}^{(k)}_I(t_k,t) V_{IJ}(t) \widetilde{U}^{(k)}_J(t_k,t)^\dagger \\
  \widetilde{U}^{(k)}_I(t_k,t) &= e^{-i \sum_{l=1}^{k-1} a_{kl} \delta_\text{t} \widetilde{G}_I^{(l)}(t)}
\end{align}
$a_{kl}$ and $c_k$ are additional Runge-Kutta coefficients.
For $k=1$, $c_1=0$ so that $t_1=t$, and $V_{IJ}^{(1)}(t_1) = V_{IJ}(t)$
  and $\widetilde{G}_I^{(1)} = G_I^{(1)} = H'_I$ from \eqnref{eq:H'} at time $t$.
We find that choosing $\widetilde{G}_I^{(l)}(t) = G_I^{(l)}(t_l)$ results in
  $O(\delta_\text{t}^2)$ simulation errors after time $t\sim 1$ using the RK4 coefficients.
We instead use
\begin{align}
 \widetilde{G}_I^{(l)}(t) = \tfrac{1}{2}  \widetilde{U}^{(l)}_I(t_l,t)^\dagger &G_I^{(l)}(t_l)
                                          \widetilde{U}^{(l)}_I(t_l,t) \nonumber\\
   + \; \tfrac{1}{2} & G_I^{(l)}(t_l)
\end{align}
  for which we observe an $O(\delta_\text{t}^3)$ simulation error after time $t \sim 1$.
For RK4, the tableau of coefficients is
\begin{equation}
  \begin{array}{c|cccc}
    c_1 &&&& \\
    c_2 & a_{21} &&& \\
    c_3 & a_{31} & a_{32} && \\
    c_4 & a_{41} & a_{42} & a_{43} & \\ \hline
        & b_1    & b_2    & b_3    & b_4
  \end{array} \;\;\scalebox{1.2}{=}\;\;
  \begin{array}{c|cccc}
    0   & &&& \\
    1/2 & 1/2 &&& \\
    1/2 &  0  & 1/2 && \\
    1   &  0  &  0  & 1 & \\ \hline
        & 1/6 & 1/3 & 1/3 & 1/6
  \end{array}
\end{equation}

\end{document}